\documentclass[12pt,preprint]{aastex}

\shorttitle{Mid-IR sources in EGS}
\shortauthors{Barmby et al.}

\newcommand{\mic}{~$\mu$m\ }
\newcommand{\mice}{~$\mu$m}
\newcommand{\mjysr}{~MJy~sr$^{-1}$}
\newcommand{\ujy}{$\mu$Jy\ }
\newcommand{\ujyy}{$\mu$Jy}

\begin{document}

\title{A catalog of mid-infrared sources in the Extended Groth Strip}

\author{
P. Barmby,\altaffilmark{1,2} 
J.-S. Huang,\altaffilmark{1} 
M.L.N. Ashby,\altaffilmark{1} 
P.R.M. Eisenhardt,\altaffilmark{3}
G.G. Fazio,\altaffilmark{1}
S.P. Willner,\altaffilmark{1} 
E.L. Wright\altaffilmark{4}
}
\altaffiltext{1}{Harvard-Smithsonian Center for Astrophysics, 60 Garden Street, Cambridge, MA 02138}
\altaffiltext{2}{Current affiliation: Department of Physics \& Astronomy, University
of Western Ontario, London, ON N6A 3K7, Canada; e-mail: pbarmby@uwo.ca}
\altaffiltext{3}{Jet Propulsion Laboratory, California Institute of Technology, Pasadena, CA 91109}
\altaffiltext{4}{UCLA Astronomy, P.O. Box 951547, Los Angeles, CA 90095-1547}

\begin{abstract}
The Extended Groth Strip (EGS) is one of the premier fields for extragalactic deep surveys. 
Deep observations of the EGS with the Infrared Array Camera (IRAC) on the Spitzer Space Telescope
cover an area of 0.38 deg$^2$ to a 50\% completeness limit of 1.5 \ujy at 3.6\mice. 
The catalog comprises 57434 objects detected at 3.6\mice, with 
84\%, 28\%, and 24\% also detected at 4.5, 5.8, and 8.0\mice.
Number counts are consistent with results from other {\it Spitzer} surveys.
Color distributions show that the EGS IRAC sources comprise a mixture of populations:
low-redshift star-forming galaxies, quiescent galaxies dominated by stellar emission
at a range of redshifts, and high redshift galaxies and AGN.
\end{abstract}


\section{Introduction}

Observations of unbiased, flux-limited galaxy samples via `blank-field' extragalactic surveys 
have been a mainstay in the field of galaxy formation and evolution for 
several decades, with the well-known Hubble Deep Field \citep{hdf96}
and Sloan Digital Sky Survey \citep{york_sdss} exemplifying two very different types of galaxy survey. 
Extending the wavelength coverage as broadly as possible has led 
to numerous changes in the understanding of how galaxies form, evolve, and interact
over cosmic time. New technologies and larger telescopes continually increase the 
volume of discovery space, making some `state-of-the-art' observations obsolete in just a few years.

The locations of extragalactic survey fields are driven by a number of considerations.
To achieve the deepest possible data, foreground 
diffuse emission and absorption should be low. Relevant properties 
include Galactic H~I column density (particularly important for X--ray observations), 
Galactic and ecliptic dust and `cirrus' foreground emission
(particularly important for infrared observations), schedulability
(for observability by space-based telescopes), and a lack of extremely bright 
foreground sources such as stars or nearby galaxies. 
There is of course a trade-off between ecliptic latitude and declination;
high-latitude fields are less easily observable from both hemispheres. 
The Extended Groth Strip (EGS), centered at $\alpha = 14^h17^m$, $\delta = +52^{\circ}30^{\prime}$,
is observable only from the north but
has excellent properties in  other categories and as such is one of a handful of
premier extragalactic survey fields. Observations of the EGS have now been made
at nearly every wavelength, with a number of projects 
(including the {\it Spitzer} Legacy project FIDEL) still ongoing.
Many of the datasets in the EGS region are described briefly by
\citet{davis07}; the same journal issue contains the results of initial studies
using the multi-wavelength dataset. As part of a public data release by the AEGIS
collaboration, this paper describes observations of the EGS 
made with the Infrared Array Camera \citep[IRAC;][]{irac} 
on the {\it Spitzer Space Telescope} \citep{sst} and presents a catalog derived from
those data.

IRAC is sensitive to radiation nearly out of reach for ground-based telescopes.
It was designed in part to study galaxies at high
redshift; its four bands at 3.6, 4.5, 5.8 and 8.0\mic probe the peak of the galaxy spectral 
energy distribution out to redshifts of $z=4$. Early results from the {\it Spitzer} mission 
\citep[e.g.,][]{irac_lbg} established that IRAC  could indeed detect $z=3$ galaxies, and
lensed sources at much higher redshifts ($z\sim 7$) have also been detected \citep{egami05}.
The population of galaxies detected with the MIPS instrument on Spitzer \citep{mips}
have been 
well-characterized \citep{pg05,elf05}, as have the IRAC sources detected in shallow
surveys such as SWIRE \citep{rr05} and the Bo\"otes field \citep{irac_shallow}.
However, the IRAC sources detected in deep observations such as those made of the
EGS (90 times the exposure time of SWIRE) or the GOODS fields (1500 times the exposure time)
have only begun to be explored \citep[e.g.,][]{pg08}. This paper presents the IRAC EGS catalog
and an examination of the source population; a companion paper (Huang et al., 2008, in prep.)
describes the use of the IRAC data in combination with optical data to derive
photometric redshifts. Other recent
papers by the IRAC team have used the EGS data to derive number counts \citep{irac_numcts},
define a class of infrared luminous Lyman-break galaxies \citep{huang05}, 
explore the mid-infrared properties of X--ray sources \citep{pb_gsx06}, 
identify mid-infrared counterparts to sub-millimeter sources \citep{ashby06}, 
investigate the contribution of mid-infrared sources to the sub-millimeter background \citep{dye06}, 
measure stellar masses for $z\sim 3$ Lyman-break galaxies \citep{dar06},
and identify 6~cm radio sources \citep{spw_vla}.

\section{Observations and data reduction}
\label{sec:obs}

The IRAC instrument was described by \citet{irac} and \citet{irac_dhb}.
The IRAC observations of the EGS were carried out as part of {\it Spitzer} 
Guaranteed Time Observing program number 8, using about 165 hours of
time contributed by {\it Spitzer} Science Working Group members G.~Fazio, G.~Rieke, and E.~Wright.
The observations were performed in two epochs, 2003 December and 2004 June/July. 
(Source variability between the two epochs is under analysis and will be discussed
in a future contribution.)
Each epoch's observations consisted of 26 Astronomical Observing Requests (AORs)
with each AOR implemented as a 2 column (across the width of the strip) by 1 row map
having 26 dithered 200~s exposures\footnote{Because of the higher background levels
in the 8.0\mic IRAC band, each 200~s exposure is implemented as four 50~s exposures.}
per map position.
The dither pattern used was the medium-scale cycling pattern, which has a median separation
of 53 pixels ($64\farcs7$) and includes half-pixel offsets.
The central positions of the maps were defined to align with the EGS position angle,
40\arcdeg\ east of north.
Since the {\it Spitzer} roll-angle is not selectable by the observer,
the correct orientation of the IRAC arrays (aligned with the EGS) was
accomplished by constraining the observation dates.
Each epoch's AORs were observed in order from south to north along
the EGS to minimize roll angle changes between adjacent AORs and
prevent gaps.
Because the array position angles changed by 180\arcdeg\ between the two epochs,
and because IRAC has two separate fields of view offset by 5\arcmin,
there are regions at the ends of the EGS with  only single-epoch coverage
in one field of view.
To summarize, the IRAC observations comprise 52 positions in a 
$2^\circ \times 10^\prime$ map, and at each position there are
52 dithered 200~s exposures at 3.6, 4.5, and 5.8~$\mu$m and
208 dithered 50~s exposures taken concurrently at 8.0~$\mu$m.  
The processed dataset includes 18924 Basic Calibrated Data (BCD) images:
only 4 of the expected frames were lost to pipeline problems.

Data processing began with the BCD images produced by 
version 14 of the Spitzer Science Center IRAC pipeline. 
Individual frames were corrected for the `muxbleed' and `pulldown' artifacts
near bright stars by fitting and subtracting a straight line
(counts as a function of pixel number) to the affected pixels.
This somewhat crude correction reduced pulldown to a level below the noise, but 
some muxbleed trails are still apparent in the output mosaics.
The known variation in point source calibration over  the IRAC arrays' 
field of view \citep{reach05} was not corrected for:
doing so would have compromised outlier detection during mosaicing and
resulted in non-flat backgrounds in the 
final mosaics which would have greatly complicated source detection.
Because the IRAC exposures are well-dithered, the magnitude of this effect
should be $<1$\% (see the IRAC Data Handbook).
IRAC photometry is known to vary slightly with source position
within a pixel, but this effect is $<2$\%
and should also average out of the highly-dithered EGS data.
The saturation limits for the mosaics are the same as those in
individual 200~s frames: 2~mJy ($m_{\rm AB}=15.7$) at 3.6 and 4.5\mice,
and 14~mJy ($m_{\rm AB}=13.5$) at 5.8 and 8.0\mice.

\begin{figure}
\plotone{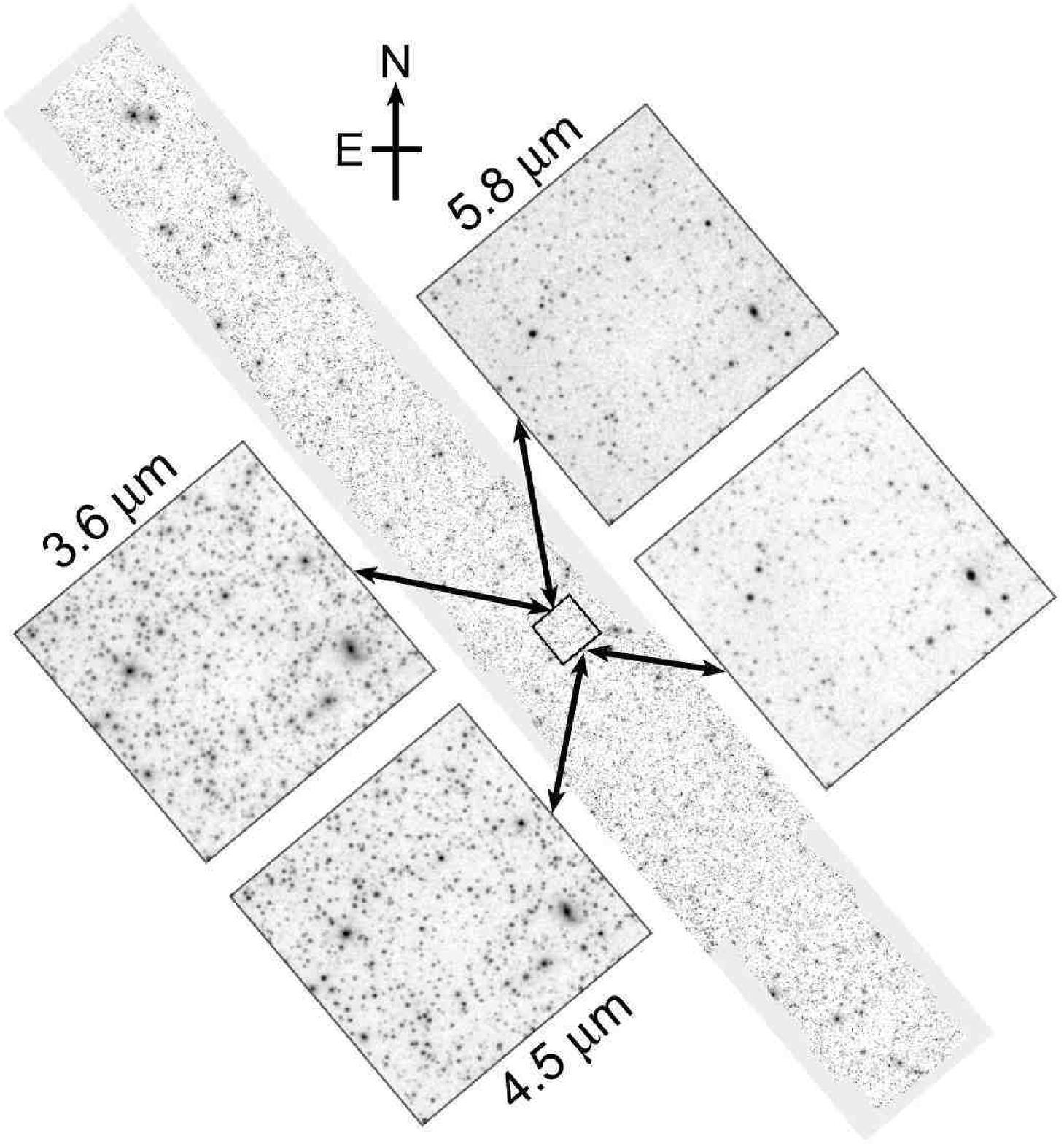}
\caption{
The Extended Groth Strip as seen by IRAC (negative image). The long image is the full
$2\fd3 \times 17\farcm3$ 3.6\mic mosaic shown with north up and east
to the left. Insets show $5\arcmin \times 5$ cutouts in each of the four bands;
the 3.6 and 4.5\mic images have 
much higher source density than the 5.8 and 8.0\mic images. 
The 7.8~arcmin$^2$ region masked due to artifacts is between the
two bright stars at the northeast end of the strip.
\label{fig:images}}
\end{figure}

Mosaicing was done using custom IDL scripts supplemented with 
procedures from the IDL Astronomy Library. 
A reference frame 
containing all the input frames in each band was constructed,
and a grid of output pixels defined. The linear dimensions of the output pixels 
were half those of the input pixels.
The individual input frames were distortion-corrected and projected 
onto the grid of output pixels using bi-linear interpolation, then the pixel stack at each
output pixel was combined by averaging with $3\sigma$-clipping.
This sigma-clipping served to reject cosmic rays, scattered light, and other image 
artifacts. Rejection of array row- and column-based artifacts was
facilitated by having the observations done at two different position angles,
but some artifacts remain in the final mosaics. The method used to remove these
from the catalog is described in \S\ref{sec:sourceid}.
After mosaics in the 4 individual IRAC bands were constructed, they
were transformed to a common pixel scale and reference frame using version 2.16.0 of 
the SWarp software written by E. Bertin, retrieved from the TERAPIX website.
The final mosaics are $2\fdg3 \times 0\fdg29$, with a pixel size ($0\farcs61$)
about half the native  IRAC pixel scale (1\farcs22).
The mosaicing and resampling  conserved surface brightness, so the mosaics,
like the input BCD images, are in units of \mjysr.
Figure~\ref{fig:images} shows a portion of the mosaics in each band.
The mosaics, coverage images, and PSF star images (see below) are available online from 
\url{http://www.cfa.harvard.edu/irac/egs/}.

\begin{figure}
\plotone{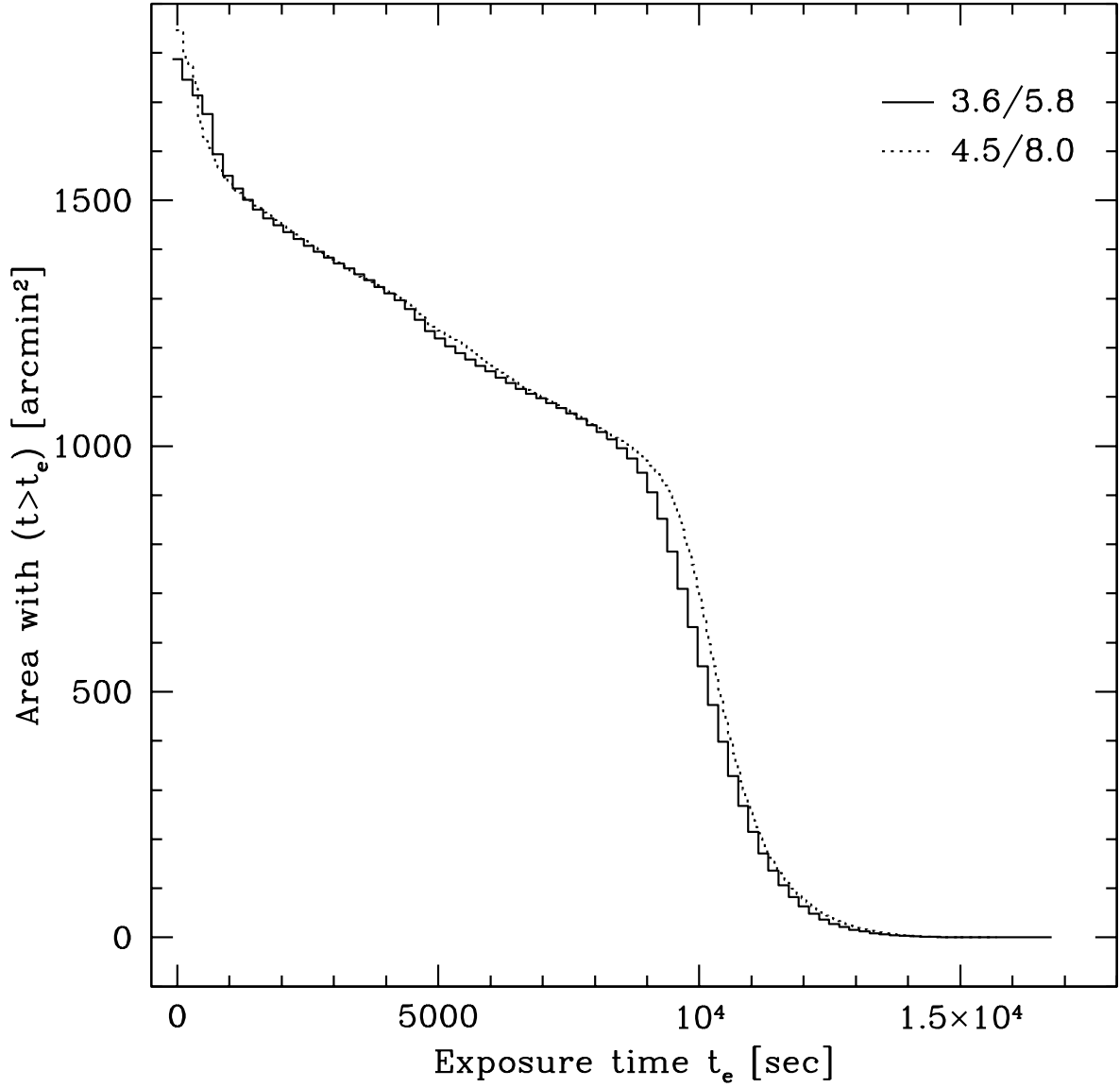}
\caption{Cumulative area coverage as a function of exposure time
for IRAC observations of the EGS. The median coverage is about 9100~s in all bands.
\label{fig:coverage}}
\end{figure}

Because of dithering the coverage varies over the mosaics;
the median coverage is about 47 frames (9100~s) in all channels.
The number of frames co-added at each position 
was recorded during processing, and a cumulative plot of the exposure time per pixel is shown
in Figure~\ref{fig:coverage}. 
Pixels with lower coverage depth are in the `crust' near the edges of the mosaic, and the
small fraction of pixels with significantly higher coverage 
are located along the center line where different map positions overlap.
Within the deep coverage area, the coverage differs slightly between the IRAC
bands due to the differing fields of view and appearance
of artifacts. (For example, cosmic ray hits affect more
pixels in the 5.8 and 8.0\mic bands, but bright-source artifacts affect more
pixels in the 3.6 and 4.5\mic bands). 
The inflection points in the coverage curves are at areas/depths of
1440~arcmin$^2$/1900~s and 930~arcmin$^2$/9100~s. About 100~arcmin$^2$ 
is covered to depths of $>11500$~s, but this deepest area is not contiguous.

Measuring the point spread functions (PSFs) of the mosaics is important for
quantifying the distribution of light within individual sources.
A `PSF star image' was constructed in each band using the SSC
prf\_estimate software (v. 030106) to combine `postage stamp' images 
of bright sources (ranging from 50 sources at 8.0\mic to 350 at 3.6\mice) 
in the mosaics. Although the PSF is known to vary over the IRAC field of view, this 
effect is smoothed over in mosaicing, and in any case the variation was
not important for our purposes. The medians and standard deviations of
photometry offsets between small measurement apertures and
the IRAC calibration photometry aperture of 12\farcs2 were computed
over the ensembles of bright sources; these are given
in Table~\ref{tab:apcor}. The aperture corrections derived here 
compare reasonably well with those found by other groups
\citep[e.g.][]{pg08,pap06,surace05}, particularly for the larger apertures
where the uncertainties are smaller.

Computing photometric uncertainties requires a good understanding of
image noise properties. Noise in the mosaics comes from several sources.  
The most fundamental is photon shot noise from the Zodiacal foreground, which for
measurements in small beams is a factor of 30 smaller than the catalog
limit at 3.6 and 4.5\mic and a factor of 10 smaller at 5.8 and
8.0~\micron.  Another noise contributor is source confusion, discussed
by \citet{dole03}.  For aperture photometry in an aperture of area
$A$ large compared to the point spread function, their equation~5
becomes
\begin{equation}
\sigma^2 = A (1.09/3600^2) B S_{lim}^2 (\gamma+2)^{-1} ,
\end{equation}
where the source density is represented by a power law
$dN/dS = B (S/S_{lim})^\gamma$~sources deg$^{-2}$ mag$^{-1}$.
Based on the source counts given in \S\ref{sec:numcts}, source confusion noise is
comparable to photon noise at 3.6 and 4.5~\micron\ and smaller than
photon noise at 5.8 and 8.0~\micron.  The present mosaics have an
additional noise source because of the method of data taking.
Observations in each location in the strip were taken at about the same
time.  Therefore any temporal drift in the IRAC zero point translates
nearly directly into a zero point shift with spatial position on the
scale of the IRAC field of view (5\arcmin).  Errors in the flat field
(gain matrix) will have a similar scale size but will only affect
bright sources.

\begin{figure}
\plotone{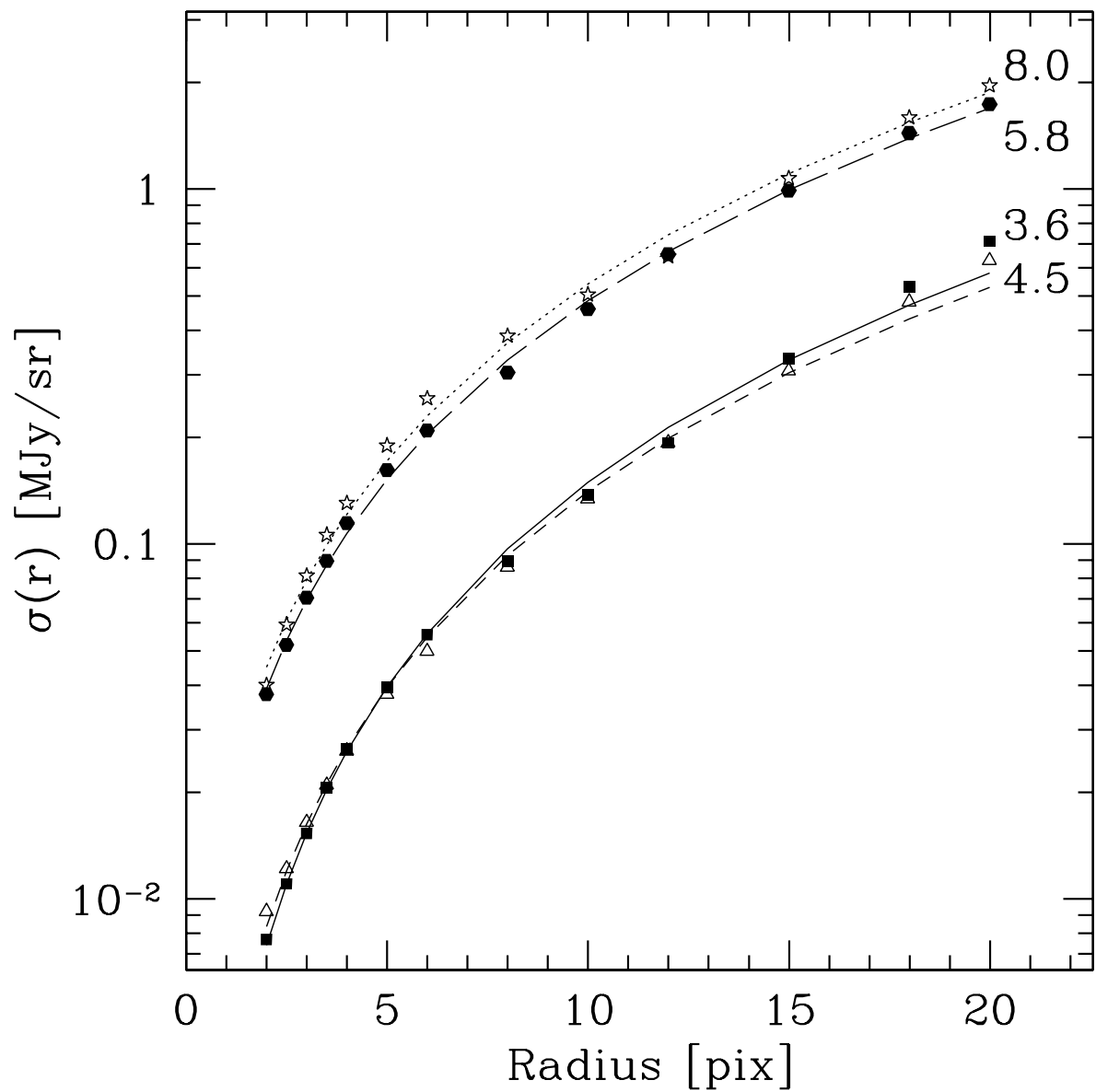}
\caption{
Standard deviations for sums of image counts (in \mjysr) measured in empty regions 
on IRAC EGS mosaics. Counts were measured in circular apertures of radius $r$ pixels.
Lines represent fits of Equation~\ref{eq:noise} to the data. Solid line is for 3.6\mic (squares);
short-dashed line for 4.5\mic (triangles); long-dashed line for 5.8\mic (hexagons);
dotted line for 8.0\mic (open stars).
\label{fig:sigma_npix}}
\end{figure}

In view of the difficulty of knowing all the noise sources, we have
adopted an empirical approach to determining the noise.  Aperture
photometry of a set of 300 locations, distributed over the field and
free of visible sources, gave a measure of variance for a range of
aperture sizes at each wavelength. The variance for aperture sizes
up to 12 pixels radius fits a function of the form
suggested by \citet{labbe03}:
\begin{equation}
\sigma(r) = \sigma_1 r (a + b r) .
\end{equation}
\label{eq:noise}
where ${\sigma}_1$ is the pixel-to-pixel RMS noise and $r$ is the aperture radius.
The data and fits are shown in Figure~\ref{fig:sigma_npix}; Table~\ref{tab:noise} 
gives the coefficients $\sigma_1$, $a$, and $b$.  The first term corresponds
to the combined effects of source confusion and Poisson noise, and the
second term corresponds to the zero point uncertainty at any location.
It is dominant for extended sources and amounts to 
0.02\mjysr at the two shorter wavelengths and 0.05\mjysr 
at the two longer wavelengths.
This noise could potentially be decreased by different reduction
techniques that force the mosaic zero point to be constant
\citep[e.g.,][]{fma00}.  For a radius of 3~pixels
(=~1\farcs8), about the smallest aperture feasible, the first term
implies noise of 0.04, 0.05, 0.5, and 0.4~$\mu$Jy in the four IRAC
bands, respectively.  For the two longer channels, this is roughly
consistent with source confusion noise and represents the approximate
limit to which an optimum technique could extract point sources.  For
the shorter two wavelengths, the empirical ``linear'' noise is much
smaller than the estimated confusion noise.  The reason is unclear,
but the most likely explanation is that the zero point uncertainty
is so large as to make it impossible to measure the empirical confusion
noise.  The total empirical noise $\sigma(r)$ in a small beam is 
consistent with expected source confusion noise.

\section{Source identification and photometry}
\label{sec:sourceid}

To construct catalogs from the IRAC EGS mosaics, we used the SExtractor
package \citep[v 2.5.0;][]{ba96} as is becoming standard in the field.
We experimented with the input parameters to achieve an acceptable
balance between completeness and reliability (as judged by eye, but see
also \S\ref{sec:complete}). The 3.6 and 4.5\mic mosaics are quite
similar to each other in degree of crowding and background level, as
are the 5.8 and 8.0\mic mosaics, but the short and long wavelength pairs
are quite different from each other, so we derived 2 sets of input
parameters for the two wavelengths regimes. The key values are given 
in Table~\ref{tab:settings}. 
Most parameter settings were close to the defaults. 
The largest differences were the choice not to filter the images prior
to detection, and setting the de-blending minimum contrast parameter
to zero for the 3.6 and 4.5\mic images: these helped to improve the de-blending of 
crowded sources, particularly at the shorter wavelengths.

The coverage images generated during mosaicing were used in two different
ways as SExtractor input. In a fairly standard procedure, the coverage maps
were used as `weight maps' for detection, such that a faint object appearing
on a deeper area of the image receives greater weight than one near the crust.
We also generated a `flag' image by combining the individual band mosaics'
coverage maps with a minimum function and
setting areas near bright stars affected by muxbleed or
pulldown to have flag values of 1. By including the flag values in the
SExtractor output, sources in regions of low coverage or near image artifacts
could be easily eliminated from the final catalog.
Mosaic regions with coverage $> 10$ images (40 images at 8.0\mice) in all bands, a total area 
of about 1362~arcmin$^2$ (0.38~deg$^2$),
were used to generate the catalog. The area within the EGS lost to artifact masking
is about 15~arcmin$^2$, half of this in a 7.8~arcmin$^2$ region
around and between the two brightest stars, centered on J2000 coordinates
$14^{\rm h}23^{\rm m}11\fs5, 53^{\rm d}34^{\rm m}02^{\rm s}$.

SExtractor was used to measure source magnitudes in a number of different ways. 
The first method is standard circular aperture photometry, in apertures
of radius 2.5, 3.5 and 5.0 pixels ($1\farcs53$, $2\farcs14$, and $3\farcs06$).%
\footnote{In apertures of this size, a significant fraction of the flux comes
from pixels only partially within the aperture. To test that SExtractor correctly
deals with these `partial pixels' we used it to perform aperture photometry on
an image with a uniform background. There were small systematic differences between
the total flux in small apertures and the expected values, but the differences were
$<1$\%, with no dependence on the aperture position relative to the pixel center.}
These magnitude were corrected to total magnitudes using point-source 
corrections derived from the mosaic PSF images
and given in Table~\ref{tab:apcor}. Also recorded were SExtractor's  AUTO
and ISOCOR magnitudes, which measure the total flux within the Kron radius
and the isophotal area above the background (with a correction for flux in the
PSF wings), respectively. 
The isophotes used for photometry were determined separately for each channel;
they correspond to the level of the detection thresholds 
above the background (given in Table~\ref{tab:settings}). 
No additional aperture corrections beyond those performed by SExtractor
were applied to the AUTO and ISOCOR magnitudes.
We did not attempt to measure magnitudes by PSF-fitting. Analysis
of very deep observations in the `IRAC Dark field' (J. Surace, priv. communication)
showed that most faint IRAC sources are slightly resolved; for these objects, 
aperture photometry is more accurate.

To compute photometric uncertainties, SExtractor assumes that the
background sky noise is Poisson and uncorrelated between adjacent pixels.
This is not the case for our re-sampled, mosaiced data, so we followed
\citet{gawiser06} in deriving a correction to the uncertainties, based on
our noise measurements in \S\ref{sec:obs}.
To correct the SExtractor flux uncertainties we apply:
\begin{equation}
\frac{\sigma_{\rm phot,corr}}{\sigma_{\rm phot,SE}} = \left(
\frac{\sigma^2(r) + \frac{F}{G}}
{\sigma^2_1 \pi r^2 + \frac{F}{G}} 
\right)^{1/2} 
\end{equation}
\label{eq:photcorr}
where $F$ is the object flux as measured in \mjysr units, $G$ is
the effective gain (electrons per image unit, see Table~\ref{tab:settings}),
and ${\sigma}^2(r)$ is computed from Eq.~\ref{eq:noise} with the coefficients for each band
given in Table~\ref{tab:noise}.
The radius $r$ is the measurement aperture for aperture magnitudes,
the Kron radius in pixels for AUTO magnitudes, and $({\rm ISOAREA}/\pi)^{1/2}$ for isophotal magnitudes.
The magnitude of the correction factor varies with aperture size and, for
objects in the number count peak, is typically about a factor of 2 for ISOCOR
and aperture magnitudes and 4 for AUTO magnitudes, which use larger apertures.

Aperture magnitudes are of course most appropriate for point sources,
and some sources in the IRAC EGS mosaics are clearly resolved.
Most of the obvious extended sources in the EGS data are bright nearby galaxies
which can be identified with the SExtractor CLASS\_STAR output parameter.
The distribution of CLASS\_STAR 
as a function of magnitude, combined with visual inspection of the images,
shows that accurate separation between resolved and unresolved objects
is possible for the 5785 sources brighter than $[3.6]_{\rm AB, auto}=20.25$. 
Brighter than this limit, 3224 sources (56\%) have CLASS\_STAR $<0.05$ and are
therefore likely to be extended. 
Comparison of the corrected aperture magnitudes to AUTO and ISOCOR values for 
these sources suggests that, as expected, the two smaller aperture magnitudes
underestimate the total flux, by 10--20\% on average. The 5-pixel-radius aperture
magnitudes are within about 5\% of the AUTO and ISOCOR measurements, as are
all of the corrected aperture magnitudes for point sources in the same magnitude range.
Most of the extended sources are relatively
small, $r<10$~arcsec. However, 13 objects are large enough 
($r_{\rm iso} = (A_{\rm iso}/\pi)^{1/2} > 12$~arcsec as measured on the 3.6\mic image)
to require the use of the `extended source calibration'.\footnote%
{See \url http://ssc.spitzer.caltech.edu/irac/calib/extcal/index.html}
Table~\ref{tab:extdcorr} gives the correction factors for each object,
derived using the measured $r_{\rm Kron}$ or $r_{\rm iso}$ in each band.
These corrections {\em have} been applied to the final catalog.

\subsection{Completeness and reliability}
\label{sec:complete}

Understanding the completeness and bias of a large survey is important
for deriving its overall statistical properties, and the standard
`artificial object' method was used to do this for the IRAC EGS catalogs.
Both point and extended sources were generated using the {\sc artdata} package
in IRAF with the mosaic point spread functions.
There are a large number of possible parameters for artificial `galaxies'
made with {\sc artdata}; we chose to use half `exponential disk'
and half `de Vaucouleurs' profiles, with axial ratios $>0.5$ and effective
radii $r_e=1$~pixel. Although this is a rather small size,
after being convolved with the PSF, the resulting sources had similar sizes to the
real objects in the images.
The artificial sources were inserted into the mosaic images, then identified and photometered
using SExtractor in the same manner as real sources. A total of 50000 artificial sources were 
inserted with power-law ($\alpha=0.3$) distributions of
magnitudes in ranges $18 <[3.6,4.5]_{AB}<26$, $17<[5.8]_{AB}<25$, 
and $16.5<[8.0]_{AB}<23.5$. The artificial sources were inserted 1500
at a time in the 5.8 and 8.0\mic mosaics, and 500 at a time in the
more-crowded 3.6 and 4.5\mic mosaics. An object was considered to be
recovered if its position was within a radius of 1.5 pixels and its magnitude within
0.5~mag of an input artificial source.
The second requirement reduces the chance that detection of a bright source 
will incorrectly be considered to be recovery of a nearby faint artificial source. 

\begin{figure}
\plotone{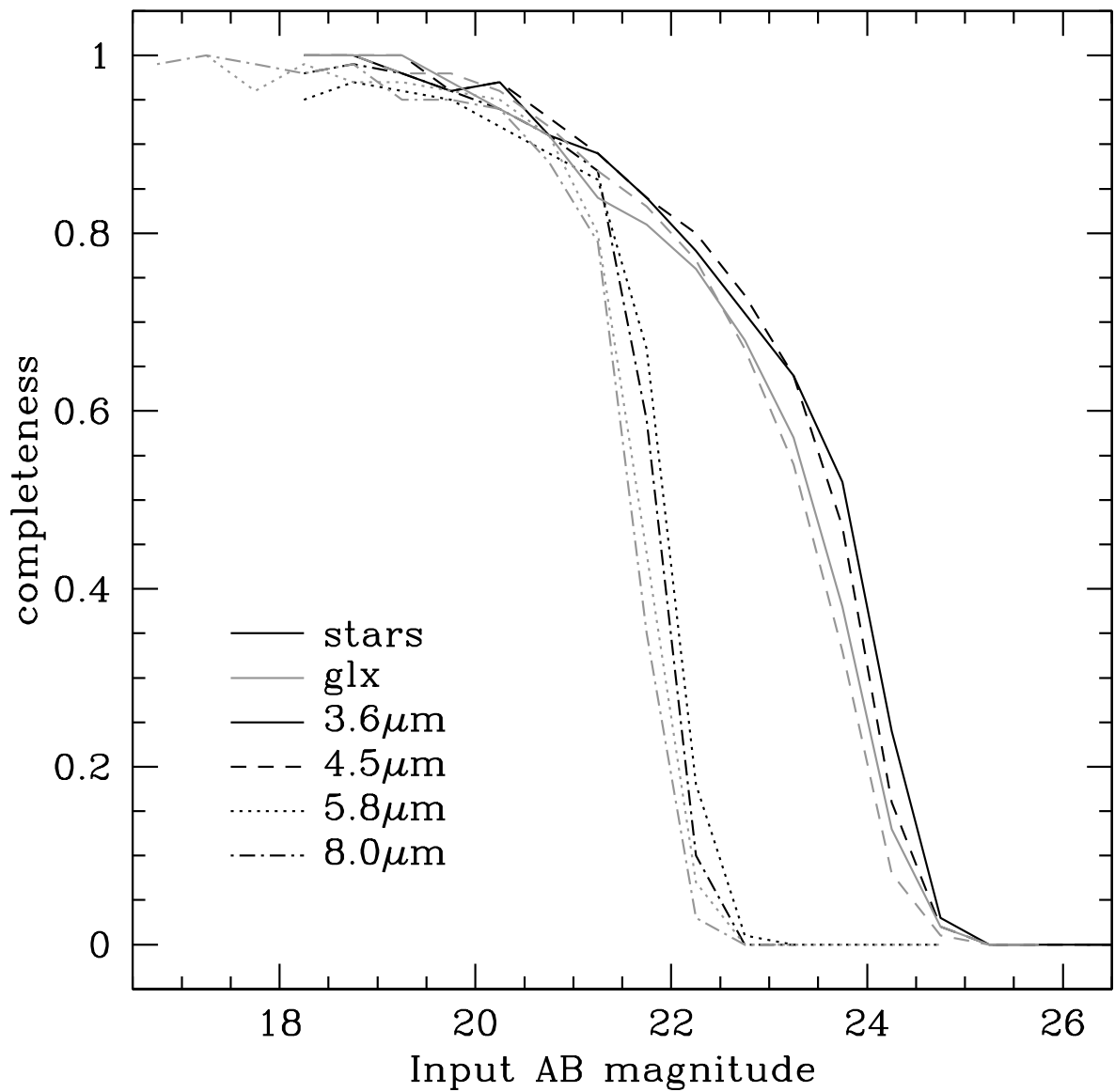}
\caption{Completeness (fraction of artificial objects recovered) 
as a function of input magnitude for IRAC observations of the EGS.
Black lines represent point sources and gray lines extended sources,
with solid, long-dashed, short-dashed, and dot-dashed lines representing
the 3.6, 4.5, 5.8, and 8.0\mic bands.
\label{fig:complete}}
\end{figure}

Completeness estimates were derived by sorting the artificial sources into bins by 
input magnitude and dividing the number of recovered sources by the number input.
The results are shown in Figure~\ref{fig:complete} and given in tabular form in Table~\ref{tab:comp}.
The completeness curves for the 3.6 and 4.5\mic bands have a somewhat shallower
fall-off than those for the 5.8 and 8.0\mic bands. 
This is likely due to the effects of crowding: some sources which are well above
the noise limit are not recovered because they fall too close to another source.
As expected, the completeness for extended sources is somewhat lower than that
for point sources. The 50\% point-source completeness limits in the 4 IRAC bands 
are $m_{\rm AB} = 23.8, 23.7, 21.9, 21.8$, or 1.1, 1.2, 6.3, and 6.9 \ujyy.
The 50\%-completeness limits for extended sources are 0.3~mag brighter,
corresponding to 1.5, 1.6, 8.3, and 9.1 \ujyy.

\begin{figure}
\plotone{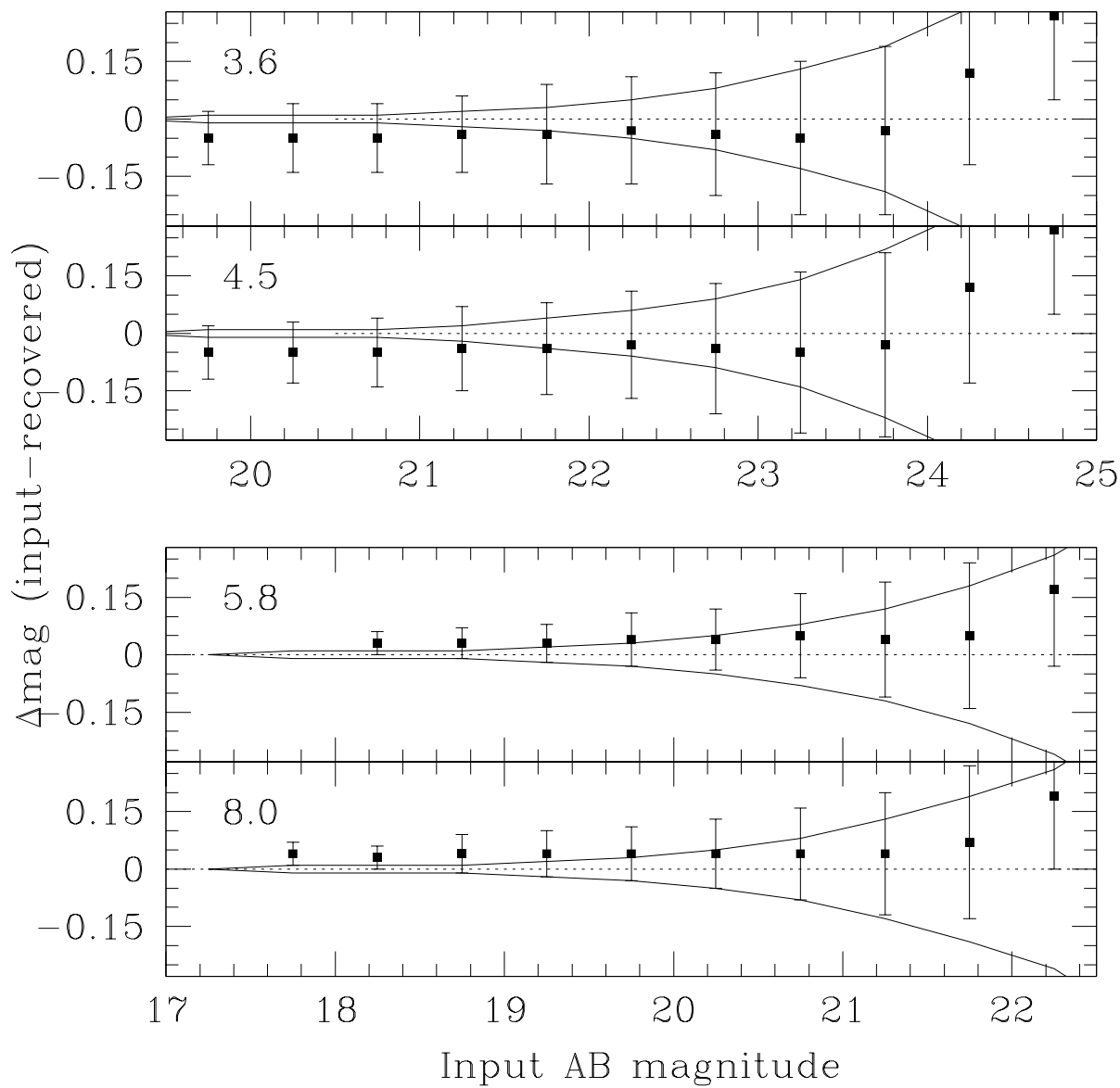}
\caption{Difference between input and recovered magnitudes for 
artificial point sources in IRAC observations of the EGS.
Recovered magnitudes are aperture magnitudes in 3.5 pixel radius apertures;
vertical error bars are the standard deviations of the magnitude offsets in each bin.
Solid lines connect the median magnitude uncertainties 
(for 3.5 pixel radius apertures) computed for the catalog
objects in the same magnitude bins.
\label{fig:mag_off_star}}
\end{figure}

\begin{figure}
\plotone{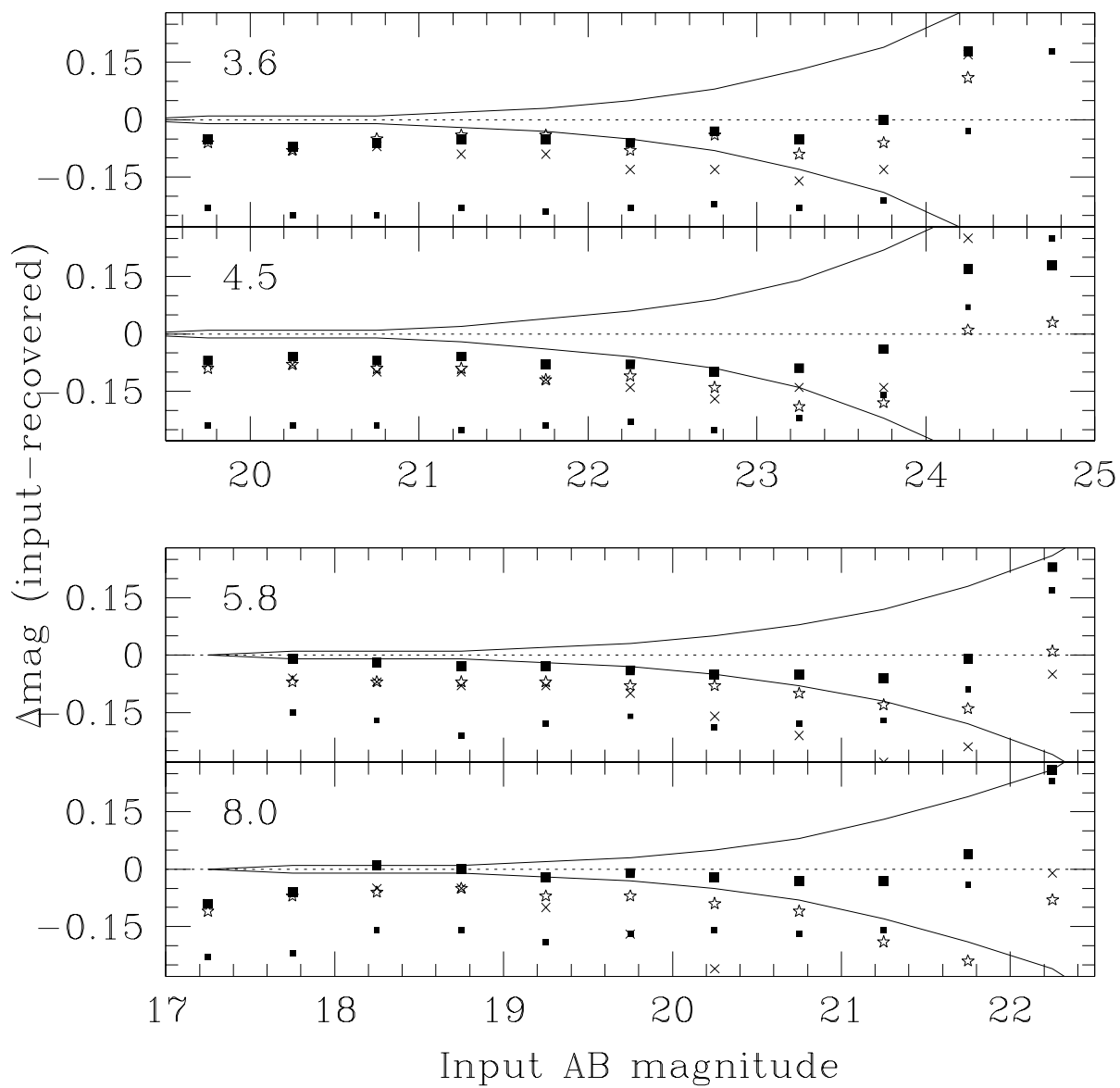}
\caption{
Difference between input and recovered magnitudes for 
artificial extended sources in IRAC observations of the EGS.
Symbols: stars: MAG\_AUTO, crosses: MAG\_ISOCOR, squares:
corrected aperture magnitudes in 2.5 pixel radius (small) or 5 pixel
radius (large) aperture. 
Solid lines are the same as in Figure~\ref{fig:mag_off_star}.
\label{fig:mag_off_glx}}
\end{figure}

The artificial source tests also permit tests of SExtractor's photometry.
Figures~\ref{fig:mag_off_star} and \ref{fig:mag_off_glx} show the 
results of sorting artificial objects into bins by input magnitude and computing
the median offsets between input and recovered magnitudes. The aperture
corrections described in \S\ref{sec:obs} were applied to the recovered aperture magnitudes.
In general, the recovered magnitudes are fainter than the input
magnitudes, but most offsets are consistent with zero within the scatter.
As expected, the small-aperture magnitudes underestimate the total fluxes of extended 
sources, but large aperture magnitudes (and to a lesser extent the isophotal and Kron magnitudes)
recover the input flux, consistent with the results in \S\ref{sec:sourceid}. 
For point sources, the standard deviations of the
magnitude offsets per bin range from about 0.01~mag for the brightest artificial
sources measured with the smallest aperture to $\sim 0.5$~mag for the faintest
artificial sources measured with MAG\_AUTO. 
These values are roughly comparable to the median photometric uncertainties
measured for the catalog objects in the same bins (shown in \ref{fig:mag_off_star} as solid lines). 
For extended sources the  bin standard deviations range from 0.08--0.6. For both point and extended sources,
the scatter between input and recovered magnitudes increases going from aperture magnitudes through 
MAG\_ISOCOR and finally to MAG\_AUTO. While these latter two magnitudes should provide
better estimates of total flux for well-resolved sources, there are relatively few 
such objects in the EGS images, and we recommend the use of aperture magnitudes
for most analyses of the catalog. In the analysis which follows in \S\ref{sec:analysis}
we use magnitudes measured in the 3.5-pixel (2\farcs1) radius aperture; this
is near the `ideal' aperture chosen by \citet{surace05} and a reasonable compromise
between the reduced contamination afforded by a smaller aperture and
the greater flux fraction of a larger one. 

To estimate the reliability of the catalog, we used the standard method of
searching for sources on a negative image. This relies on the assumption 
that the noise is symmetric with respect to the background. Using the same SExtractor
parameters described in \S\ref{sec:sourceid}, 640 sources were detected on the
3.6\mic image in the (coverage $> 10$) region used to generate the catalog.  
The probability of spurious sources being detected at the same position
in more than one band
is proportional to the source density multiplied by the image and matching disk
areas, and is $\lesssim 10^{-4}$ for the source density found in the 3.6\mic negative image.
Negative versions of the 4.5, 5.8 and 8.0\mic mosaics were created and analyzed
in the same method as for the final catalog (by association
with a source in the negative 3.6\mic image; see \S\ref{sec:bandmatch}).
No matched sources were found in the longer-wavelength negative mosaics at the same significance
levels used for the real catalogs.
Therefore the only possible spurious sources are those detected only at 3.6\mice.
There are about 9400 such sources, with an estimated spurious fraction of
$640/9400 = 6.8$\%.  
This gives an overall spurious fraction for the full 3.6\mic selected
catalog of 1.1\%, or a reliability of 99\%.

\subsection{Astrometry}
\label{sec:astrom}

\begin{figure}
\plotone{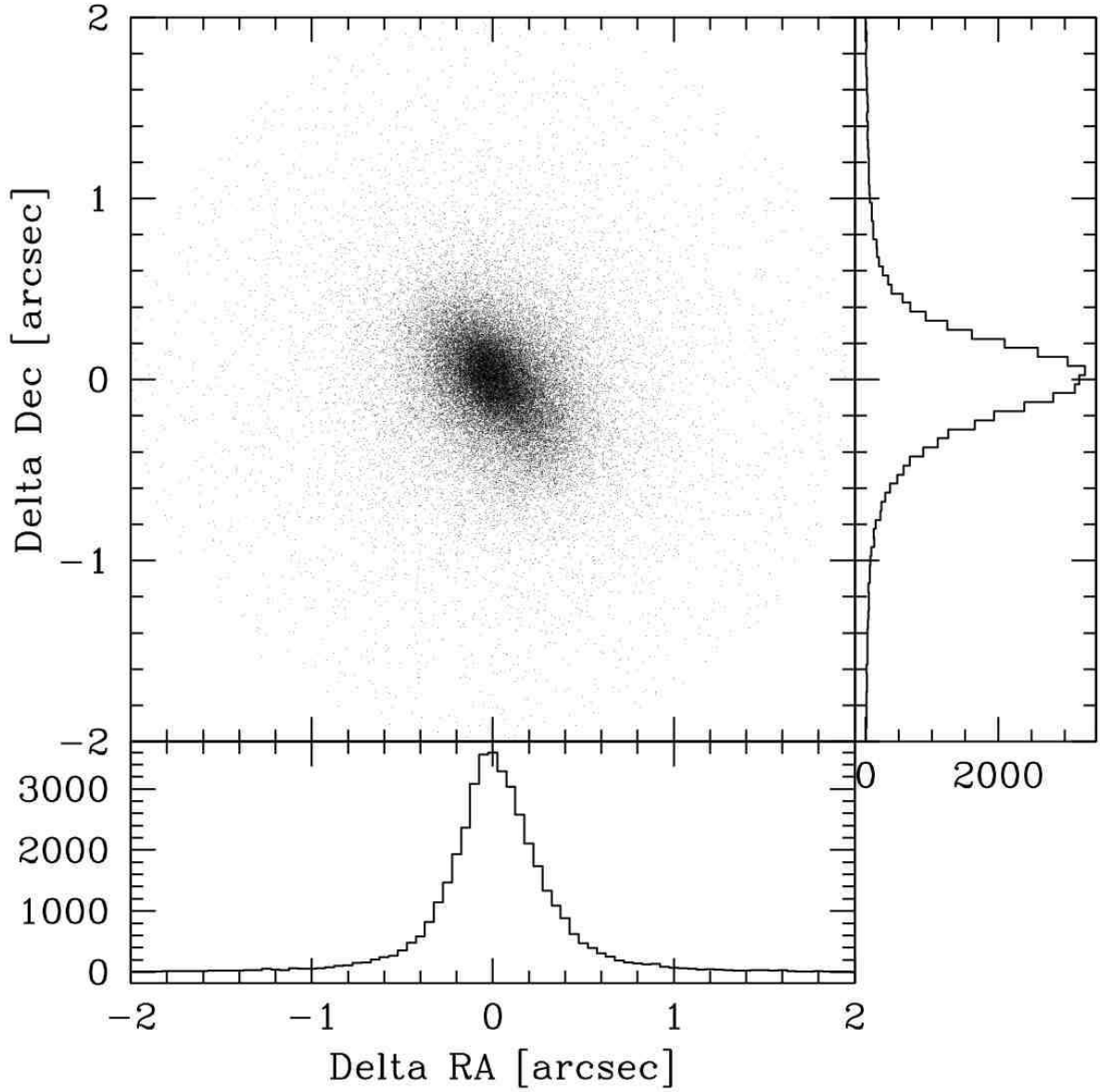}
\caption{Astrometric offsets between IRAC source positions and those of sources
in the DEEP2 photometric catalog, matched with a positional tolerance of
2\farcs0.
\label{fig:del_ra_dec}}
\end{figure}

\begin{figure}
\plotone{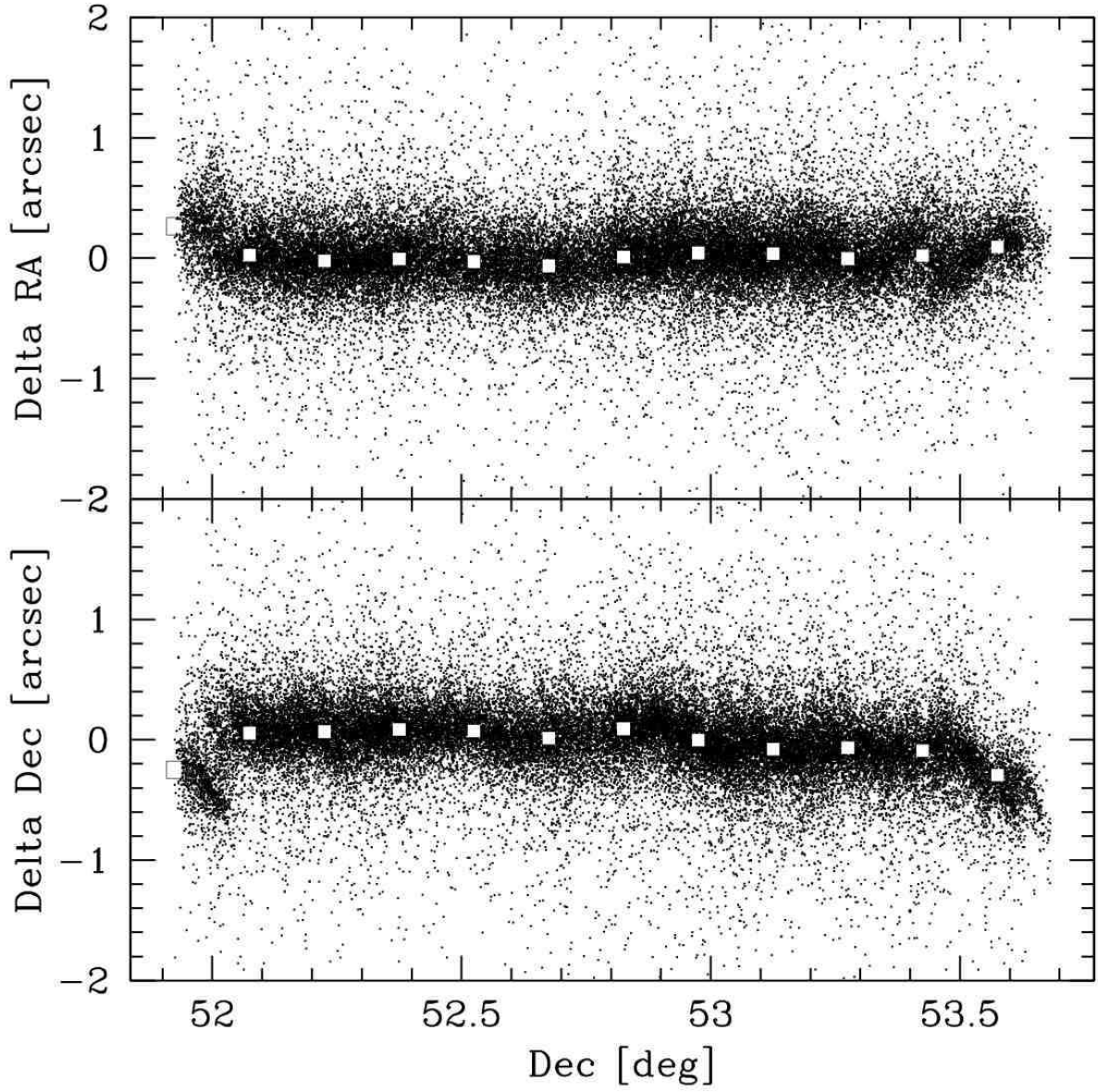}
\caption{
Astrometric offsets between IRAC and DEEP2 sources, as a function of
position. Solid squares are median values in 0\fd1 bins.
\label{fig:astrom_trend}}
\end{figure}

The precision and accuracy of positional measurements is an important quality
in a large astronomical catalog. The quality of the astrometry in the IRAC mosaics 
is determined by both the world coordinate systems for the individual BCD images
and the accuracy with which they are combined. To assess the astrometric quality
of the IRAC catalog, we matched 3.6\mic sources within a 2\farcs0 radius to optical sources from the
DEEP2 photometric catalog \citep{deep2opt}, which is tied to the SDSS coordinate frame.
Figures~\ref{fig:del_ra_dec} and \ref{fig:astrom_trend}
show the results. The accuracy of the IRAC astrometry is very high overall: the median
offset is  $0\farcs012$ ($-0\farcs004$) in RA (declination). The precision, as indicated 
by the standard deviations of the offsets (both 0\farcs37), is consistent
with expectations from the size of the IRAC PSF and pixels. There are larger offsets
in both RA and Dec at the northern and southern ends of the EGS: these correspond to regions
where the IRAC data were taken at only one epoch (see \S~\ref{sec:obs}).
Evidently averaging two array position angles along the center of the EGS
improved small errors in astrometry. To maintain consistency between the
catalog and released mosaic images, we have not adjusted the positions of
sources in the regions near the ends of the strip to make the median offsets equal to zero
(they are still consistent with zero within our quoted precision). 
Catalog users wishing to adjust the astrometry for these sources should 
add $(0\farcs2,-0\farcs2)$ to the coordinates of objects with $\delta<52\fdg025$ and
$(0\farcs03,-0\farcs1)$ to the coordinates of objects with $\delta>53\fdg525$.

\subsection{Band-matching}
\label{sec:bandmatch}
The combination of measurements in the 4 IRAC bands was done using 
SExtractor's `association' mechanism: the 3.6\mic catalog was used as the
master catalog, with sources in the other 3 bands associated by pixel position.
We chose this method rather than `dual-image' mode (in which source and
aperture positions are derived from a master image and used identically
on other images) because there were small ($<2$ pixel, or 1\farcs2) but noticeable shifts 
between the mosaics in different bands, particularly at the ends of the EGS. 
The offset appears to be between the two IRAC fields-of-view (3.6/5.8 and 4.5/8.0),
suggesting some relation to the mapping strategy used, although its exact
cause is unclear.
The offsets would have been problematic for dual-image mode, but the shifts were
small enough that objects were matched between catalogs without difficulty.
Requiring a 3.6\mic detection does not unduly bias the catalog: 
less than a few hundred objects are detected at 4.5, 5.8 or 8.0\mice without a 
corresponding 3.6\mic detection. All of these objects are faint;
many ``8.0\mice-only objects'' are in fact detections of the Galactic cirrus
emission at the southeast end of the EGS while others are the result of
differences in de-blending between different bands.

The IRAC EGS catalog is presented in Table~\ref{tab:cat}.\footnote{%
Also available at \url{http://www.cfa.harvard.edu/irac/egs/}.}
This is a 3.6\mice-selected catalog, so all objects are detected
in this band. Objects undetected in the other bands have all parameters
listed as zero.
The aperture corrections given in Table~\ref{tab:apcor}
have been applied to all aperture magnitudes.
As discussed in \S\ref{sec:sourceid}, only the area of sky with
exposure time $>2000$~s in all 4 IRAC bands was used to generate the catalog.
Positions reported are as measured on the 3.6\mic image (see \S\ref{sec:astrom}
for discussion of astrometric accuracy).
The magnitude uncertainties given are statistical and do not include 
the systematic calibration uncertainty \citep[2\%;][]{reach05}. 
Saturation limits are (see \S\ref{sec:obs}) 
$m_{\rm AB} = 15.7, 15.7, 13.5, 13.5$ or 2, 2, 14 and 14 mJy
in the 4 IRAC bands.
The columns of Table~\ref{tab:cat} are described in Table~\ref{tab:catcol}.
The first 7 columns are given only once per object, and the remaining columns 
once per band per object.

Because the 3.6 and 4.5\mic bands are more sensitive than the 5.8 and 8.0\mic 
bands, many sources are detected in only the two short-wavelength images.
The 3.6~$\mu$m selected catalog contains 57434 objects, with 48066, 16286, and 13556
detected at 4.5, 5.8 and 8\mic (detection fractions of 84, 28, and 24\%).
While the 3.6 and 4.5\mic images have similar sensitivities, some 
of the faintest objects are too blue to be detected at 4.5\mice:
objects at the 3.6\mic detection
limit will only have a 4.5\mic detection if they have 
$[3.6]-[4.5] > -2.5\log[f_{\lim}(3.6)/f_{\lim}(4.5)] \sim +0.1$. 
In the interests of releasing as complete a catalog as possible,
we have included all SExtractor detections in Table~\ref{tab:cat}. The
signal-to-noise of these detections, as measured by photometric
uncertainty within the 3.5-pixel radius aperture, goes down
to about $S/N \sim 2$. A less-complete but more-reliable catalog
is also available through the website listed above, in which
we have included only objects detected with signal-to-noise $\geq 5$.
This catalog reaches just below the 50\% completeness levels
and includes 44772, 38017, 13486 and 11546 sources in the 4 IRAC bands.

Confusion is significant in the two shorter-wavelength images of the EGS.
The number of beams per source, based on a beam area $\Omega = \pi \sigma^2$
\citep[$\sigma={\rm FWHM/2.35}$;][]{hogg01},%
\footnote{Some authors use a definition of $\Omega$ which is twice as large, which
reduces the number of beams per source by a factor of 2.}
is about 28 at 3.6\mice, 35 at 4.5\mice, and $\sim 97$ at 5.8 and 8.0\mice. 
Another measure of confusion is provided by matching IRAC sources with those from
a catalog at higher resolution. Such a catalog is available from the Hubble
Space Telescope Advanced Camera for Surveys (ACS) observations of the central
70\farcm5 $\times$ 10\farcm1 of the EGS: there are about $8\times 10^4$ ACS sources
(to $I_{\rm AB}=28.1$) in this area and about $3.1 \times 10^4$ IRAC 3.6\mic sources. With a match radius
of 2\farcs0, about 93\% of 3.6\mic sources were matched to an ACS source.
The IRAC sources without ACS counterparts comprise 2 groups: stars which 
show diffraction spikes on the ACS image but are not included in the ACS catalog ($\sim 10$\%),
and sources which are undetected on the optical image. About half of 
the ACS-undetected sources are relatively bright ($[3.6]_{\rm AB}\lesssim 21$);
these interesting sources will be followed up in a future contribution.
About 30\% of the matched IRAC sources had two or more ACS sources within 2\farcs0
and roughly 7\% had three or more ACS sources within this radius.
Although SExtractor attempts to correct for flux from neighboring objects when doing
photometry, up to one-third of IRAC sources may have their photometry affected at some level 
by confusion.

\section{Analysis}
\label{sec:analysis}

\subsection{Number counts}
\label{sec:numcts}
\begin{figure}
\plotone{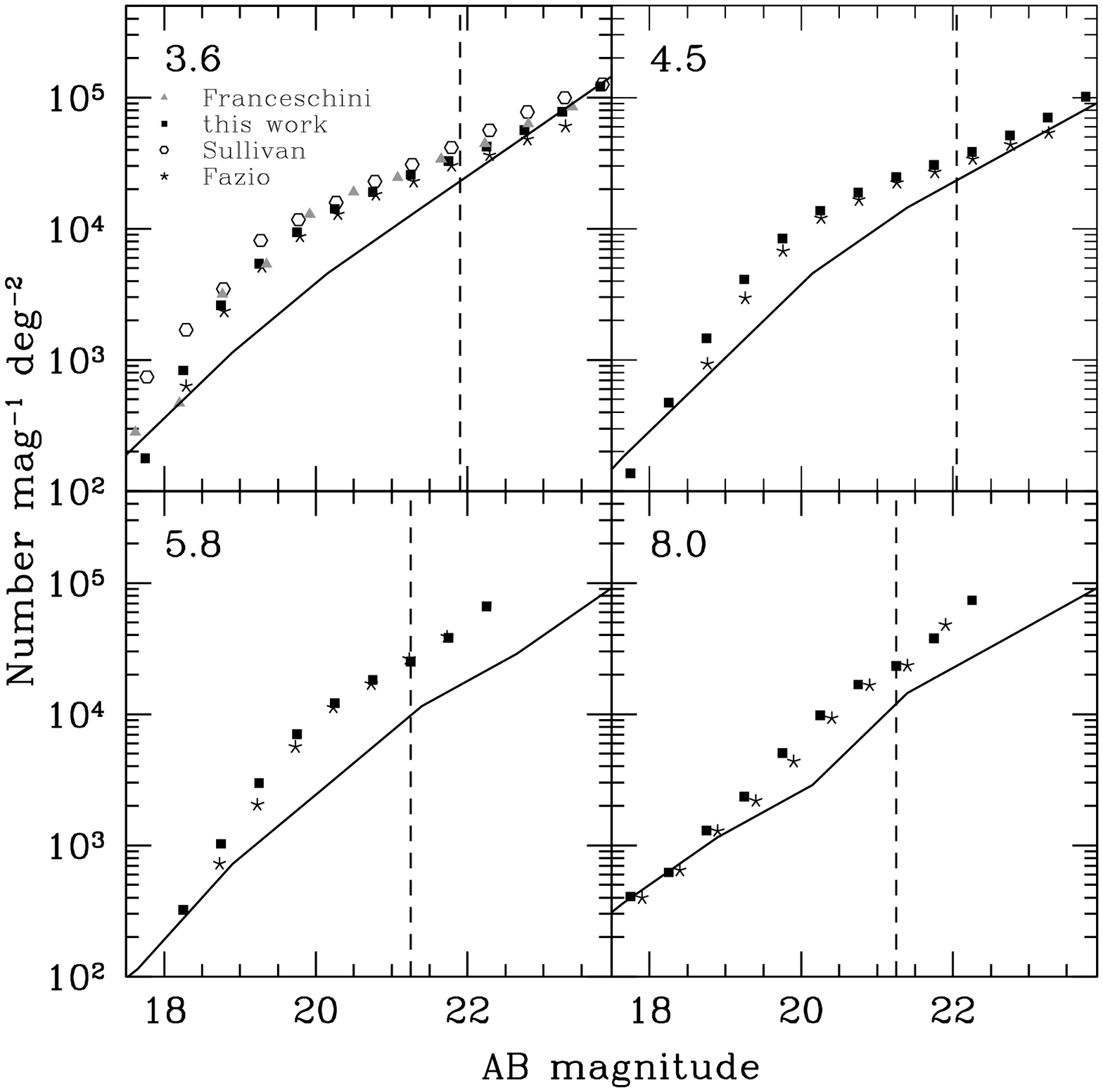}
\caption{Differential number counts derived from IRAC surveys.
Squares: this work (with model star counts  subtracted), asterisks: EGS number counts from \citet{irac_numcts},
open circles: number counts from \citet{sull07},
triangles: number counts from \citet{fra06}.
Solid lines: models (`total counts') from \citet{lacey07}.
All counts are corrected for incompleteness; vertical dashed lines show the 80\% completeness
limit of the present IRAC EGS catalog.
\label{fig:numcts}}
\end{figure}

A fundamental property of any astronomical catalog is the distribution of
sources as a function of flux. To compare our catalog with other recent
work, we derived number counts of galaxies using the SExtractor aperture magnitudes
in the 2\farcs1-radius aperture. The counts have been
corrected for incompleteness using the results of \S\ref{sec:complete}.
The star count model for the EGS given in \citet{irac_numcts}
was subtracted from the raw number counts; no other attempt was made to separate 
stars and galaxies.
Figure~\ref{fig:numcts} shows the number counts derived from the EGS data
and compares them to other recent measurements in the IRAC bands \citep{irac_numcts,fra06,sull07}
and the models of \citet{lacey07}.
There is excellent agreement with the results of \citet{irac_numcts}, as
expected since the data and analysis methods used are very similar.
Our number counts are reasonably consistent with previous results, except at 
the faintest magnitudes where our incompleteness may be underestimated. 
The \citet{lacey07} models produce the correct general trends but are offset
from the data by up to a factor of 2, a feature also apparent in their Figure~1.
\citet{lacey07} did not consider this offset serious since their models had not been
tuned to match the {\it Spitzer} data.

\subsection{Color distributions}

Galaxy colors in the IRAC bands are affected by a number of components:
the Rayleigh-Jeans tail of emission from starlight,
PAH emission, the redshifted 1.6\mic stellar opacity minimum,
and (often red power-law) emission from an AGN. Determining the
dominant source of emission for IRAC sources is complicated by
the lack of redshift information for many sources; IRAC's sensitivity
allows it to detect galaxies in the `redshift desert' where optical
spectroscopic redshifts are not easy to obtain. But a general picture
of the IRAC source can be derived by examination of color distributions
and comparison with models and other surveys.
In the following analysis, all colors are measured using aperture magnitudes
(including aperture corrections) in the aperture with radius 3.5 pixels (2\farcs1).

\begin{figure}
\plotone{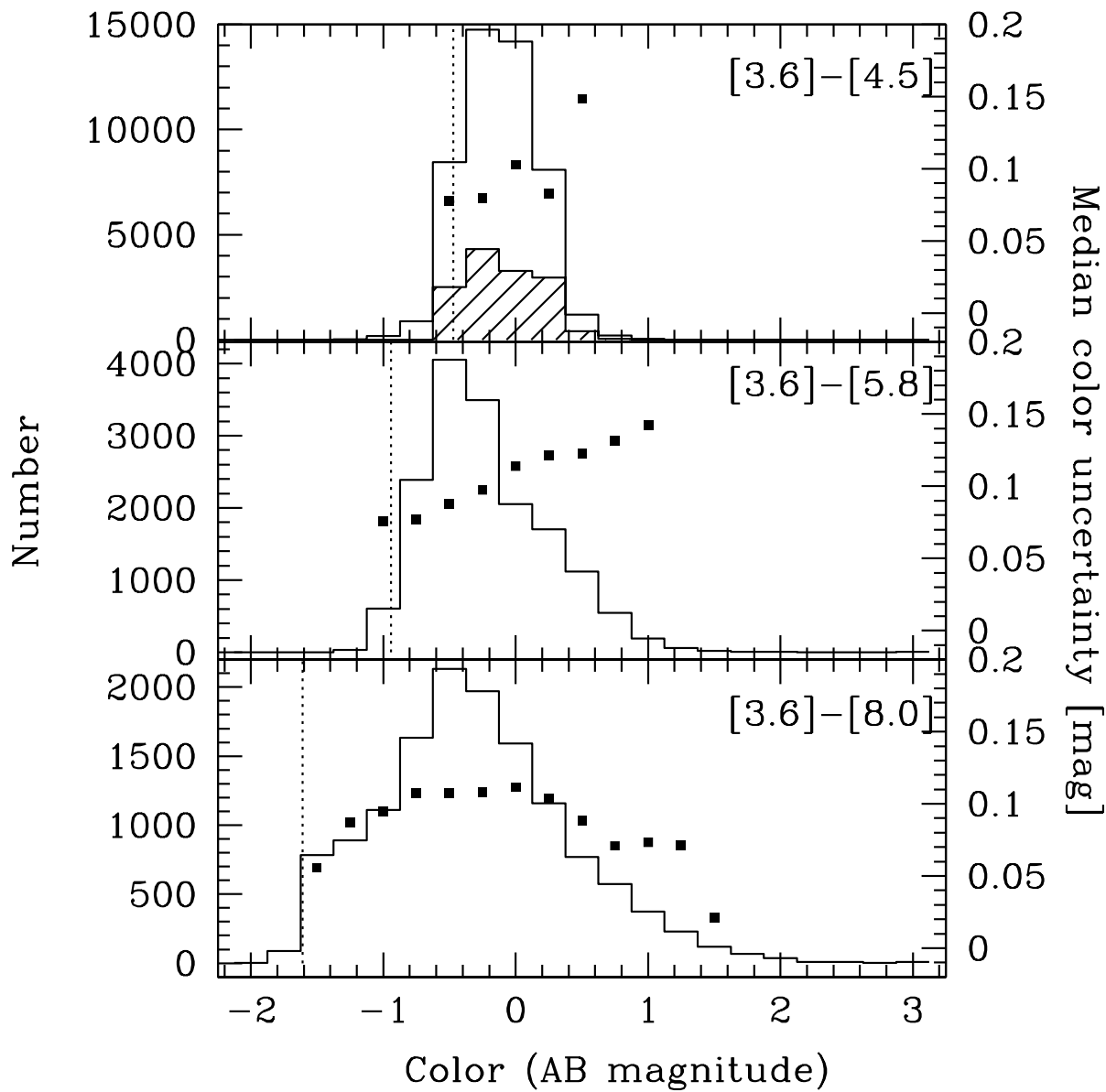}
\caption{Distribution of IRAC colors for sources in the EGS catalog.
The vertical lines denote the AB magnitude colors corresponding
to Vega magnitudes of zero (the color expected for starlight).
The shaded histogram in the top panel shows the distribution of $[3.6]-[4.5]$
magnitudes for sources with an 8.0\mic detection.
Filled boxes indicate median color uncertainties in each color bin, according to the
scale on the right-hand side of the plot.
All colors in this and following plots are based on corrected aperture magnitudes in a 
3.5-pixel (2\farcs1) radius aperture.
\label{fig:colordist}}
\end{figure}

\begin{figure}
\plotone{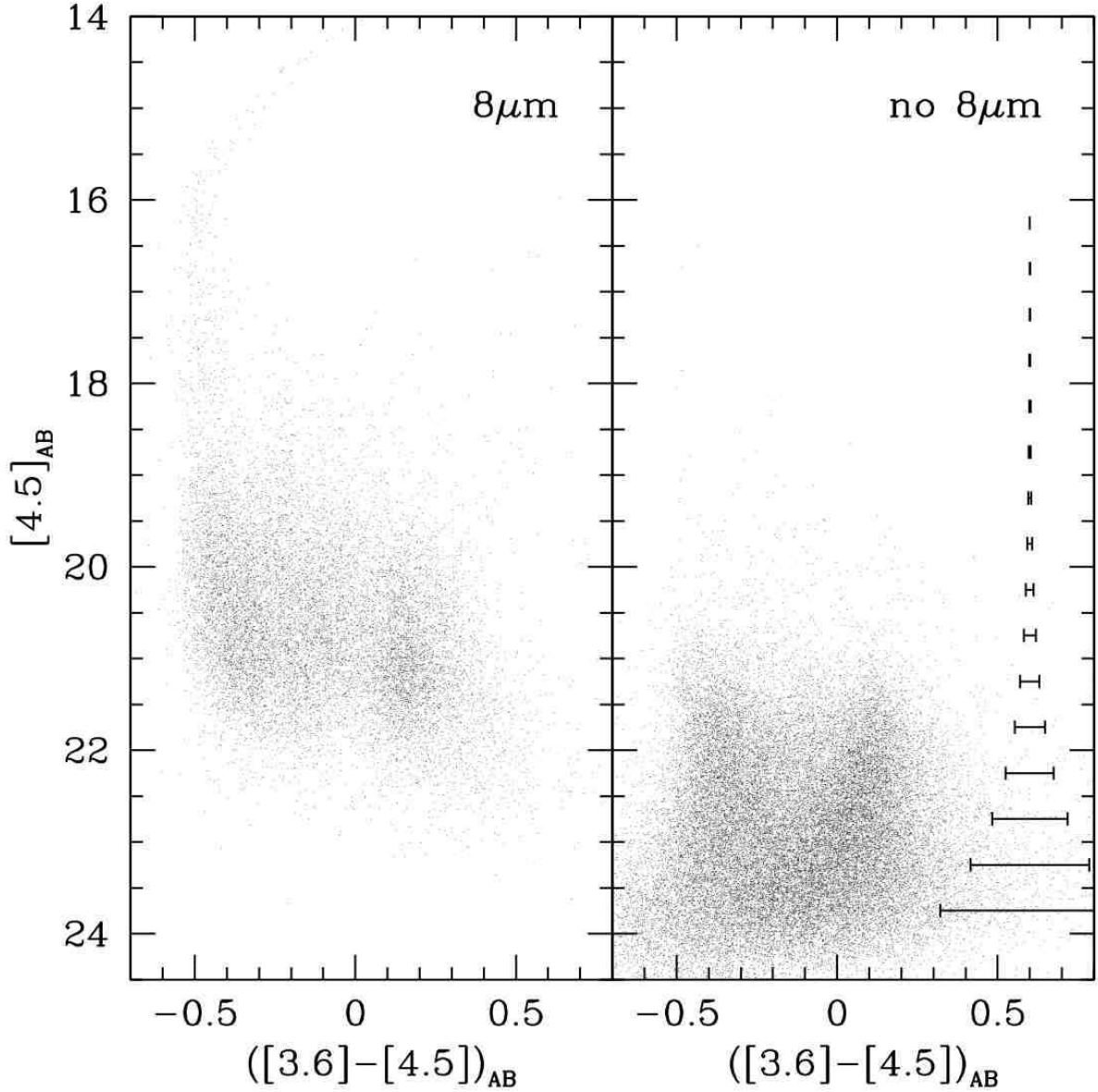}
\caption{IRAC color-magnitude diagrams $[3.6]-[4.5]$ versus $[4.5]$,
using aperture magnitudes measured in 2\farcs1 radius apertures. 
Left: 13538 sources with an 8.0\mic detection.
Right: 34538 sources without an 8.0\mic detection.
Horizontal error bars in the right panel indicate median color
uncertainties in magnitude bins.
\label{fig:cmd}}
\end{figure}

Figure~\ref{fig:colordist} shows the distribution of IRAC source colors relative
to the 3.6\mic band. 
As expected, few sources are bluer than unreddened stars, although PAH emission
in the 3.6\mic band and CO absorption in the 4.5\mic band can cause some
bluer colors. The $[3.6]-[4.5]$ color distribution is relatively narrow
and is similar for sources with and without 8.0\mic detections.
The colors involving the two longer wavelength bands show much more dispersion,
presumably because they depends on the variable strengths of the PAH features
moving through the bands with redshift \citep[see also Figure 6 of][]{huang07}.
Figure~\ref{fig:cmd} shows a color-magnitude diagram for sources with
and without 8.0\mic detections; the latter are simply fainter.
The bright, blue objects in the left-hand panel are stars; the red measured colors
for the brightest objects are due to saturation in the 3.6\mic photometry.

\begin{figure}
\plotone{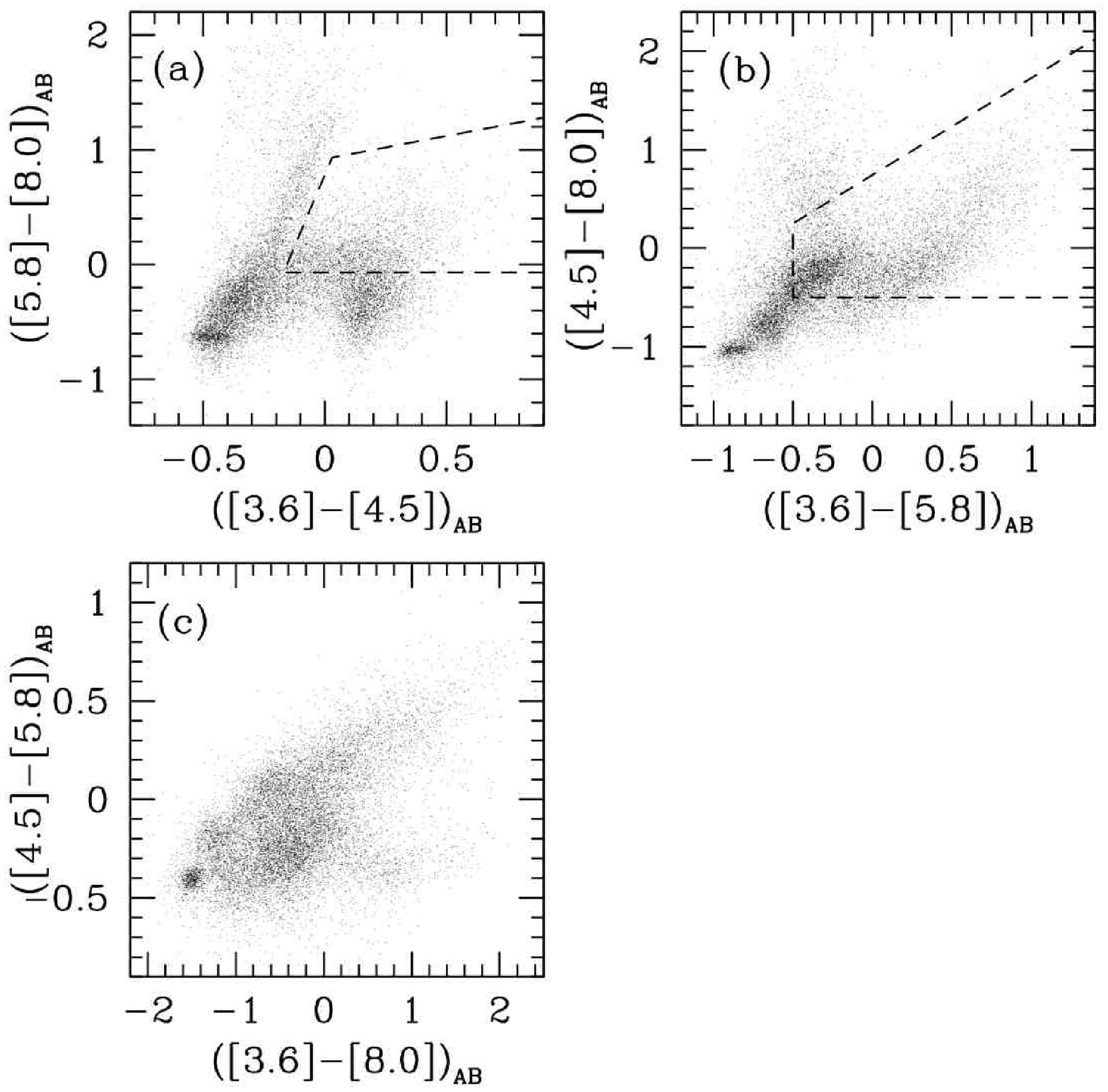}
\caption{Two-color diagrams using 4 bands for sources in the EGS catalog,
using aperture magnitudes measured in 2\farcs1 radius apertures. 
Only sources with four-band detections are plotted. 
Panel (a) corresponds to the color space used by \citet{stern05} and 
panel (b) to that used by \citet{lacy04} and \citet{sajina05}; dashed
lines show the outline of their `AGN wedges'.
The tight condensation of points at blue colors corresponds to galaxies
dominated by stellar emission while the vertical or diagonal branches
contain low-redshift galaxies and mixtures of high-redshift galaxies and AGN; see text for details.
\label{fig:4color}}
\end{figure}

The 4 IRAC bands can be combined in a number of ways to make two-color diagrams.
Different authors plot these in different ways: as flux ratios, colors in the
Vega system, and colors in the AB system. 
We have plotted all such diagrams in the AB system, which has the
advantage that different combinations of colors can be easily compared, 
but the disadvantage of complicating comparisons to previous work.
Figure~\ref{fig:4color} show two-color diagrams using 
the 3 possible combinations of all 4 IRAC bands.  The three diagrams have some common features:
a relatively tight distribution of sources with the bluest colors, and two branches at redder colors.
The blue sources are particularly well-separated in Figure~\ref{fig:4color}c
and are presumably dominated by stellar emission.
In the models of \citet{sajina05}, using the color space of Figure~\ref{fig:4color}b,
the vertical branch is dominated by
low-redshift galaxies with PAH emission, and the redder diagonal branch 
(which dominates the EGS distribution) is expected to be some mixture of AGN and high-redshift galaxies. 
Comparing Figure~\ref{fig:4color}a to Figure~1 of \citet{stern05}, the EGS 
catalog appears to contains fewer low-redshift, PAH-dominated galaxies (upper left)
but more sources in the `AGN wedge' (centre right), and the location
expected for high-redshift normal galaxies (lower right).
This is consistent with the fainter flux limit of the EGS observations:
these should contain more high-redshift galaxies
than the IRAC Shallow Survey sources with optical spectroscopy
plotted by \citet{stern05}, and at $z \gtrsim 2$,
star-forming galaxies begin to have similar observed colors to AGN
\citep{pb_gsx06}.
Similar conclusions can be drawn from
comparison of Figure~\ref{fig:4color} with Figure~1a of \citet{davoodi06}: 
as expected, the EGS has a lower proportion of low-redshift galaxies 
compared to the shallower but wider SWIRE survey.

\begin{figure}
\plotone{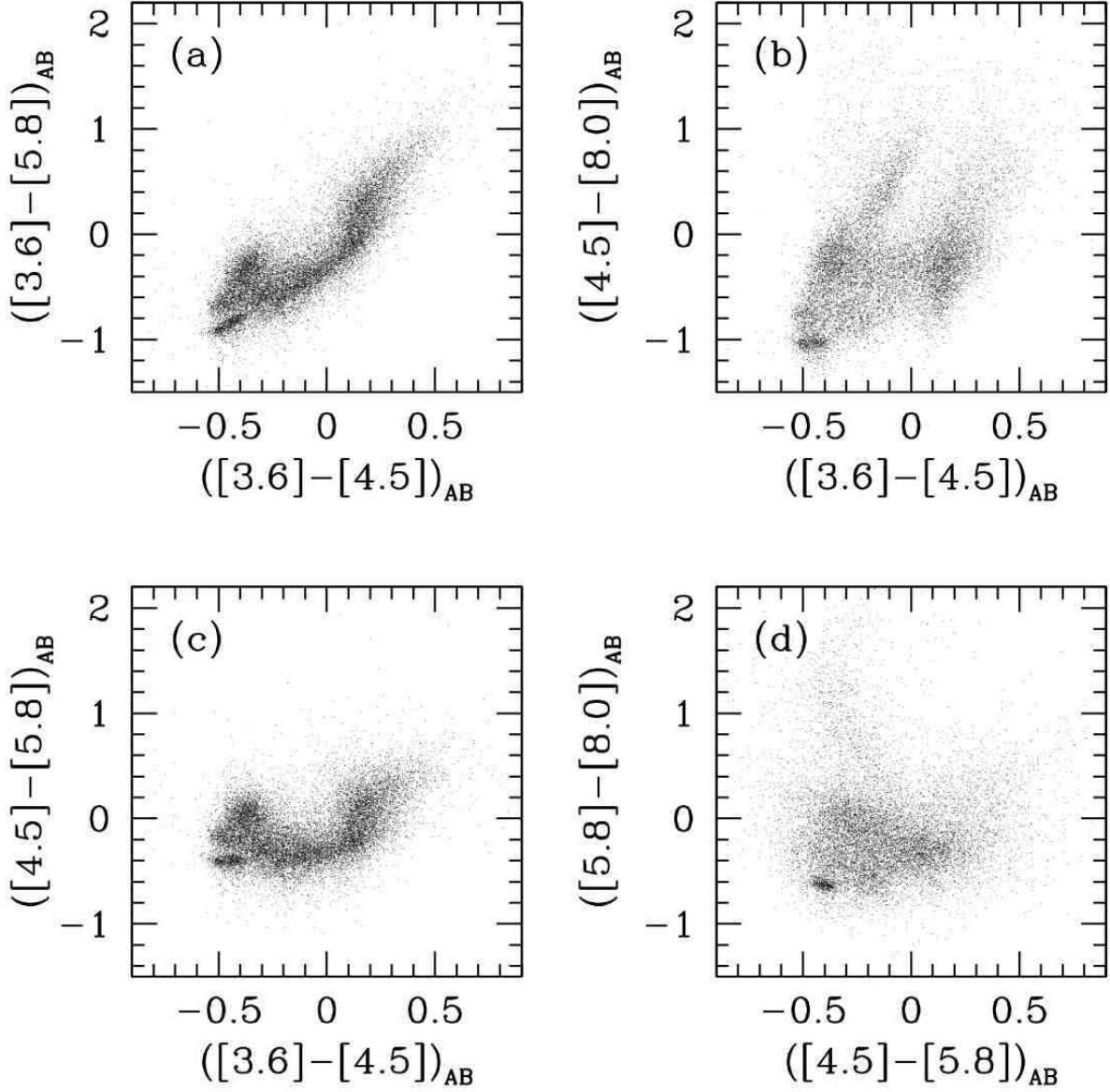}
\caption{Two-color diagrams using 3 bands for sources in the EGS catalog,
using aperture magnitudes measured in 2\farcs1 radius apertures. 
Only sources with detections in all 3 relevant bands are plotted.
See text for interpretation of color distributions.
\label{fig:3color}}
\end{figure}

There are many more combinations of three IRAC bands than can be conveniently plotted;
Figure~\ref{fig:3color} shows a few. The color space shown in Figure~\ref{fig:3color}a
does not appear to be useful for separating different galaxy types; the sources all
lie roughly along a single axis. Figure~\ref{fig:3color}b is quite similar to 
Figure~\ref{fig:4color}a, which might suggest that the 5.8\mic band does not provide much additional
information over the combination of the other 3 bands. However, Figure~\ref{fig:3color}c 
shows that the use of the three shortest bands works well to identify red sources.
This color space was used by \citet[Figure 4]{hatz05} to suggest a color criterion
for type 1 AGN. However, \citet{pb_gsx06} found that only about 30\% of 
X--ray selected AGN in the EGS fell into their selection region.
\citet[Figure 1c]{davoodi06} suggest that objects red in both $[3.6]-[4.5]$ and $[4.5]-[5.8]$ 
are a mixture of AGN and star-forming galaxies. The EGS contains a greater proportion
of these objects than the SWIRE survey, as shown above. Figure~\ref{fig:3color}d
is the same color space plotted in Figure~1b of \citet{davoodi06}; as seen there, the
omission of the 3.6\mic band appears to decrease the separation between the various galaxy types.

\section{Summary}

Observations of a 0.38~deg$^2$ area in the Extended Groth Strip using the 
Infrared Array Camera (IRAC) on the Spitzer Space Telescope detected
tens of thousands of mid-infrared sources. The 3.6 \mice-selected catalog
presented here includes 57434 sources, of which most are detected
at 4.5\mic and roughly one-quarter are detected at 5.8\mic and 8.0\mice.
Number counts of sources are consistent with previous observations
and marginally consistent with recent models. As expected, color distributions differ from
those of shallower surveys by including a greater fraction of potential high-redshift sources.
Future projects possible with this catalog include determination of
photometric redshifts, galaxy stellar mass and luminosity functions, and
mid-infrared characterization of populations such as luminous infrared galaxies 
and AGN.

\acknowledgments
We thank the referee for a thorough review which pointed out several important
issues.
This work is based on observations made with the {\it Spitzer Space Telescope},
which is operated by the Jet Propulsion Laboratory, California Institute of 
Technology under a contract with NASA. Support for this work was provided by NASA 
through an award issued by JPL/Caltech.

Facilities: \facility{Spitzer(IRAC)}

\bibliographystyle{aas}

\clearpage

\clearpage

\begin{deluxetable}{llllllll}
\tabletypesize{\scriptsize}
\tablecaption{Aperture corrections for Extended Groth Strip IRAC mosaics\label{tab:apcor}}
\tablewidth{0pt}
\tablehead{
\colhead{Band} & \multicolumn{6}{c}{Aperture radius}\\
\colhead{} & \colhead{2.5 pix ($1\farcs53$)} & \colhead{3.5 pix ($2\farcs14$)} & \colhead{5.0 pix ($3\farcs06$)}
& \colhead{2.45 pix ($1\farcs5$)} & \colhead{3.3 pix ($2\farcs0$)} & \colhead{4.9 pix ($3\farcs0$)}}
\startdata
3.6 & $-0.61\pm0.05$ & $-0.31\pm0.03$ &$-0.16\pm0.03$&$-0.63\pm0.05$ & $-0.35\pm0.04$ &$-0.16\pm0.03$\\
4.5 & $-0.62\pm0.04$ & $-0.33\pm0.03$ &$-0.15\pm0.03$&$-0.62\pm0.04$ & $-0.37\pm0.04$ &$-0.16\pm0.03$\\
5.8 & $-0.83\pm0.04$ & $-0.49\pm0.04$ &$-0.23\pm0.03$&$-0.83\pm0.04$ & $-0.54\pm0.04$ &$-0.24\pm0.03$\\
8.0 & $-0.95\pm0.02$ & $-0.62\pm0.02$ &$-0.37\pm0.02$&$-0.95\pm0.02$ & $-0.66\pm0.02$ &$-0.38\pm0.02$
\tablecomments{Corrections to be added to aperture magnitudes to convert them to 
total magnitudes.}
\enddata
\end{deluxetable}

\begin{deluxetable}{lllll}
\tabletypesize{\scriptsize}
\tablecaption{Background noise fits for IRAC mosaics\label{tab:noise}}
\tablewidth{0pt}
\tablehead{
\colhead{Band}  & \colhead{$a$} & \colhead{$b$} & \colhead{${\sigma}_1$}}
\startdata
3.6 & $0.54$ & $0.85$ & $1.66\times10^{-3}$ \\
4.5 & $0.92$ & $0.66$ & $1.87\times10^{-3}$ \\
5.8 & $2.19$ & $0.59$ & $6.65\times10^{-3}$ \\
8.0 & $1.96$ & $0.71$ & $6.31\times10^{-3}$ \\
\enddata
\tablecomments{Fits are to Equation~\ref{eq:noise}, with terms defined in \S\ref{sec:obs}.}
\end{deluxetable}

\begin{deluxetable}{lll}
\tabletypesize{\scriptsize}
\tablecaption{Parameter settings for SExtractor\label{tab:settings}}
\tablewidth{0pt}
\tablehead{
\colhead{Parameter} & \colhead{3.6/4.5} & \colhead{5.8/8.0}}
\startdata
DETECT\_MINAREA [pixel] & 5 & 5 \\
DETECT\_THRESH & 1.5 & 3\\
FILTER & N & N \\
DEBLEND\_NTHRESH & 64 & 64\\
DEBLEND\_MINCONT & 0 & 0.005\\ 
SEEING\_FWHM [arcsec] & 1.8 & 2.0 \\
GAIN & 3,2.65 $\times10^5$ & 6.28,18.5 $\times10^4$\\
BACK\_SIZE [pixel] & 200 & 200\\
BACK\_FILTERSIZE & 3 & 3 \\
BACKPHOTO\_TYPE & LOCAL & LOCAL\\
BACKPHOTO\_THICK & 24 & 24 \\
WEIGHT\_TYPE & MAP\_WEIGHT & MAP\_WEIGHT\\
\enddata
\end{deluxetable}

\begin{deluxetable}{lllllllll}
\tablecolumns{9}
\tabletypesize{\scriptsize}
\rotate
\tablecaption{Photometry corrections for individual extended sources\label{tab:extdcorr}}
\tablewidth{0pt}
\tablehead{
\colhead{EGSIRAC} & \multicolumn{4}{c}{AUTO magnitudes} & \multicolumn{4}{c}{ISO magnitudes}\\
\colhead{} &
\colhead{$[3.6]$} & \colhead{$[4.5]$} & \colhead{$[5.8]$} & \colhead{$[8.0]$} &
\colhead{$[3.6]$} & \colhead{$[4.5]$} & \colhead{$[5.8]$} & \colhead{$[8.0]$}
}
\startdata
EGSIRAC J141503.63+520434.1 & 0.07 & 0.03 & 0.15 & 0.23 & 0.07 & 0.03 & 0.10 & 0.17\\
EGSIRAC J141503.93+520909.6 & 0.07 & 0.04 & 0.16 & 0.25 & 0.07 & 0.02 & 0.11 & 0.24\\
EGSIRAC J141545.95+521328.0 & 0.05 & 0.01 & 0.09 & 0.20 & 0.06 & 0.02 & 0.07 & 0.21\\
EGSIRAC J141600.38+520617.5 & 0.08 & 0.04 & 0.19 & 0.26 & 0.06 & 0.02 & 0.10 & 0.20\\
EGSIRAC J141607.60+520810.7 & 0.06 & 0.02 & 0.09 & 0.17 & 0.06 & 0.02 & 0.06 & 0.11\\
EGSIRAC J141612.11+520936.8 & 0.06 & 0.02 & 0.14 & 0.21 & 0.06 & 0.02 & 0.08 & 0.18\\
EGSIRAC J141747.26+524102.8 & 0.07 & 0.02 & 0.13 & 0.21 & 0.06 & 0.01 & 0.05 & 0.20\\
EGSIRAC J141807.07+524150.1 & 0.07 & 0.03 & 0.13 & 0.19 & 0.06 & 0.02 & 0.09 & 0.18\\
EGSIRAC J141910.27+525151.1 & 0.05 & 0.01 & 0.11 & 0.23 & 0.06 & 0.02 & 0.08 & 0.20\\
EGSIRAC J142012.48+530729.7 & 0.06 & 0.02 & 0.12 & 0.22 & 0.06 & 0.02 & 0.06 & 0.21\\
EGSIRAC J142054.17+530705.7 & 0.07 & 0.03 & 0.19 & 0.25 & 0.07 & 0.03 & 0.15 & 0.20\\
EGSIRAC J142149.83+532005.2 & 0.05 & 0.01 & 0.07 & 0.14 & 0.06 & 0.02 & 0.04 & 0.08\\
EGSIRAC J142156.23+532601.7 & 0.05 & 0.01 & 0.07 & 0.16 & 0.06 & 0.01 & 0.03 & 0.13
\enddata
\tablecomments{
Corrections are in magnitudes to be added to the AUTO and ISOCOR magnitudes, and
have already been applied to the magnitudes in Table~\ref{tab:cat}.
Values are derived from extended source correction formula at
{\url http://ssc.spitzer.caltech.edu/irac/calib/extcal/index.html}.
}
\end{deluxetable}

\begin{deluxetable}{lllllllll}
\tablecolumns{9}
\tabletypesize{\scriptsize}
\rotate
\tablecaption{Completeness estimates for IRAC EGS catalog\label{tab:comp}}
\tablewidth{0pt}
\tablehead{
\colhead{AB magnitude} & \multicolumn{4}{c}{point source} & \multicolumn{4}{c}{extended source}\\
\colhead{} &
\colhead{$[3.6]$} & \colhead{$[4.5]$} & \colhead{$[5.8]$} & \colhead{$[8.0]$} &
\colhead{$[3.6]$} & \colhead{$[4.5]$} & \colhead{$[5.8]$} & \colhead{$[8.0]$}
}
\startdata
17.25&\nodata &\nodata &\nodata &\nodata &\nodata &\nodata &  1.00 & 1.00\\
17.75&\nodata &\nodata &\nodata & \nodata  &\nodata &\nodata &  0.96 & 0.99\\
18.25& 1.00& 1.00&  0.95& 0.98&  1.00&1.00&  0.99 & 0.98\\
18.75& 1.00& 1.00&  0.97& 0.99&  1.00&1.00&  0.97 & 0.99\\
19.25& 0.98& 1.00&  0.96& 0.98&  1.00&0.98&  0.97 & 0.95\\
19.75& 0.96& 0.96&  0.95& 0.96&  0.97&0.98&  0.96 & 0.95\\
20.25& 0.97& 0.97&  0.92& 0.94&  0.94&0.96&  0.95 & 0.94\\
20.75& 0.91& 0.93&  0.89& 0.91&  0.91&0.92&  0.91 & 0.88\\
21.25& 0.89& 0.89&  0.86& 0.87&  0.84&0.87&  0.80 & 0.79\\
21.75& 0.84& 0.84&  0.67& 0.59&  0.81&0.83&  0.44 & 0.35\\
22.25& 0.78& 0.80&  0.18& 0.10&  0.76&0.77&  0.07 & 0.03\\
22.75& 0.71& 0.73&  0.01& 0.00&  0.68&0.67&  0.00 & 0.00\\
23.25& 0.64& 0.64&  0.00& 0.00&  0.57&0.54&  0.00 & 0.00\\
23.75& 0.52& 0.47&  0.00& \nodata  &  0.38&0.33&  0.00 & \nodata\\
24.25& 0.24& 0.16&  0.00& \nodata  &  0.13&0.08&  0.00 & \nodata\\
24.75& 0.03& 0.02&  0.00& \nodata  &  0.02&0.01& \nodata &\nodata\\
25.25& 0.00& 0.00& \nodata& \nodata &  0.00&0.00& \nodata &\nodata
\enddata
\end{deluxetable}

\begin{deluxetable}{lrrrrrrrrrrr}
\tabletypesize{\scriptsize}
\rotate
\tablecaption{Extended Groth Strip 3.6\mice-selected catalog\label{tab:cat}}
\tablewidth{0pt}
\tablehead{
\colhead{EGSIRAC} &\colhead{RA} & \colhead{Dec.} & \colhead{Class} & \colhead{Flags} & \colhead{Cov} & \colhead{$r_{1/2}$} &
\colhead{} & \colhead{} & \colhead{} & \colhead{} & \colhead{}\\
\colhead{}
& \colhead{$X_i$}
& \colhead{$Y_i$}
& \colhead{$A_i$}
& \colhead{$r_{k,i}$}
& \colhead{$a_i$}
& \colhead{$b_i$}
& \colhead{$\theta_i$}
& \colhead{$m_{\rm AU,i}$}
& \colhead{$m_{\rm ISO,i}$}
& \colhead{$m_{\rm AP,i}$}
& \colhead{$\sigma(m_{\rm AP,i}$)}
}
\startdata
J141405.74+520024.2 & 213.523955 &  52.00674 &  0.212 & 0 & 11 &  1.774 &\nodata &\nodata   &\nodata  &\nodata &\nodata \\     
\nodata &13140.84& 1386.28& 20 & 4.18& 1.15&1.04 &$-56.21$  &$23.12 \pm 0.22$ & $23.06 \pm 0.15$ & 22.82  22.86  22.98 &0.12  0.18 0.34\\
\nodata &13141.13& 1385.58& 16 & 5.59& 1.19&0.86 &$-60.09$  & $23.25\pm 0.32$  & $23.23\pm  0.16$ & 22.94  22.97 23.12 & 0.15 0.21 0.37\\
\nodata &    0.00 &   0.00&   0& 0.00& 0.00&0.00 &$ 0.00 $& $  0.00\pm0.00$ &$  0.00\pm 0.00$&  0.00  0.00  0.00&  0.00  0.00  0.00\\
\nodata &    0.00 &   0.00&   0& 0.00& 0.00&0.00 &$ 0.00 $& $  0.00\pm0.00$ &$  0.00\pm 0.00$&  0.00  0.00  0.00&  0.00  0.00  0.00\\
J141406.12+520018.1 &  213.525528 &      52.005052 &    0.169 &    0  & 11 & 1.537 &\nodata &\nodata   &\nodata  &\nodata &\nodata \\     
\nodata & 13145.00 & 1375.56&  11&  5.20&  1.04& 0.96 &74.63&  $23.96\pm 0.70$&$ 23.67\pm 0.21$& 23.50  23.60  23.93 & 0.24 0.36 0.84\\
\nodata &    0.00 &   0.00&   0& 0.00& 0.00 &0.00 &$ 0.00 $& $  0.00\pm0.00$ &$  0.00\pm 0.00$&  0.00  0.00  0.00&  0.00  0.00  0.00\\
\nodata &    0.00 &   0.00&   0& 0.00& 0.00 &0.00 &$ 0.00 $& $  0.00\pm0.00$ &$  0.00\pm 0.00$&  0.00  0.00  0.00&  0.00  0.00  0.00\\
\nodata &    0.00 &   0.00&   0& 0.00& 0.00 &0.00 &$ 0.00 $& $  0.00\pm0.00$ &$  0.00\pm 0.00$&  0.00  0.00  0.00&  0.00  0.00  0.00\\
J141406.46+515947.4 & 213.526918 &  51.99651 &  0.012 &   3 & 11 &  2.268 &\nodata &\nodata &\nodata  &\nodata  &\nodata \\     
\nodata & 13180.00 & 1339.90& 57 & 16.49 & 1.52&1.43& $-30.59$& $22.33\pm 1.11$&  $22.01\pm0.17$& 22.48  22.42 22.32 &0.09 0.12  0.12 \\
\nodata & 13181.47 & 1338.56& 48 & 14.38 & 1.70&1.43& 67.45& $22.29\pm 0.83$&  $22.12\pm0.15$& 22.54  22.44 22.34 &0.10 0.12  0.17\\
\nodata &    0.00 &   0.00&   0& 0.00& 0.00 &0.00 &$ 0.00 $& $  0.00\pm0.00$ &$  0.00\pm 0.00$&  0.00  0.00  0.00&  0.00  0.00  0.00\\
\nodata &    0.00 &   0.00&   0& 0.00& 0.00 &0.00 &$ 0.00 $& $  0.00\pm0.00$ &$  0.00\pm 0.00$&  0.00  0.00  0.00&  0.00  0.00  0.00
\enddata
\tablecomments{
The complete version of this table is in the electronic edition of the Journal.
The printed version is only a sample.}
\end{deluxetable}

\begin{deluxetable}{lll}
\tablecolumns{9}
\tabletypesize{\scriptsize}
\rotate
\tablecaption{Column descriptions for IRAC EGS catalog\label{tab:catcol}}
\tablewidth{0pt}
\tablehead{
\colhead{Column} & \colhead{Description} & \colhead{units}}
\startdata
ID &  format EGSIRAC Jhhmmss.ss+ddmmss.s & \nodata\\
ALPHAWIN\_J2000 &  Right ascension in epoch J2000\tablenotemark{a} & degrees \\
DELTAWIN\_J2000 &  Declination in epoch J2000 & degrees\\
CLASS\_STAR &  SExtractor classification in 3.6\mic image, from 0 (non-stellar) to 1 (stellar) & \nodata\\
FLAGS &       SExtractor FLAGS in 3.6\mic image, range 0--3\tablenotemark{b} & \nodata\\
COVERAGE &    minimum coverage in 4 bands at object location\tablenotemark{c} & \nodata\\
FLUX\_RADIUS &      radius containing 50\% of enclosed flux at 3.6\mic & pixel \\
XWIN\_IMAGE\_i &   object barycenter in band i& pixel \\
YWIN\_IMAGE\_i &   object barycenter & pixel \\
ISOAREA\_IMAGE\_i &   isophotal area above detection threshold & pixel \\
KRON\_RADIUS\_i  &     Kron radius & pixel\tablenotemark{d} \\
AWIN\_IMAGE\_i &          semi-major axis  & pixel \\
BWIN\_IMAGE\_i &          semi-minor axis  & pixel \\
THETAWIN\_J2000\_i &      position angle, east of north & degrees \\
MAG\_AUTO\_i &         Kron magnitude & AB mag \\ 
MAGERR\_AUTO\_i &      Kron magnitude uncertainty & AB mag \\ 
MAG\_ISOCOR\_i &          magnitude in isophote above detection threshold & AB mag \\ 	    
MAGERR\_ISOCOR\_i &       isophotal magnitude uncertainty & AB mag \\ 
MAG\_APER\_i &         aperture magnitudes in 2.5,3.5, and 5-pixel radii & AB mag \\ 	    
MAGERR\_APER\_i &      aperture magnitude uncertainties & AB mag 
\enddata
\tablenotetext{a}{SExtractor's  `windowed' parameters for image location and shape (e.g.,
ALPHAWIN\_IMAGE, AWIN\_IMAGE) are used because the extensive comparison by \citet{becker08} 
showed that these were superior to the older `isophotal' 
measurements (e.g., ALPHA\_J2000, A\_IMAGE).}
\tablenotetext{b}{FLAGS is the bitwise sum of values 1 (object has near neighbors or bad pixels) or 2 (object
was originally blended with another one).}
\tablenotetext{c}{Minimum was computed as $\min(C(3.6),C(4.5),C(5.8),C(8.0)/4)$ where $C(\lambda)$ 
is the number of frames combined in band $\lambda$ at the object location.}
\tablenotetext{d}{SExtractor outputs this parameter in units of semi-major axis; the value
given here is multiplied by A\_IMAGE to convert to pixels.}
\end{deluxetable}

\end{document}